\documentclass[sigconf, nonacm]{acmart}
\AtBeginDocument{%
  }

\usepackage{amsmath}

\usepackage{graphicx}
\usepackage{booktabs}
\usepackage{array}
\usepackage{tabularx}
\usepackage{xcolor}
\usepackage{colortbl}

\usepackage[ruled,vlined,linesnumbered]{algorithm2e}

\usepackage[verbatim]{minted}

\hypersetup{
    colorlinks=true,
    linkcolor=black,
    citecolor=black,
    urlcolor=black
}

\setminted{
    fontsize=\footnotesize,
    linenos=false,
    autogobble=true,
    breaklines=true,
    breakanywhere=true
}

\SetAlgoCaptionSeparator{.}
\SetAlCapSkip{0.5em}
\SetAlgoInsideSkip{smallskip}

\newcolumntype{L}[1]{%
    >{\raggedright\arraybackslash}p{#1}%
}
\newcolumntype{C}[1]{%
    >{\centering\arraybackslash}p{#1}%
}

\begin{document}

\title{Functionally Grading the Slicing Process by Compiling Design Intent into Slicer Projects}

\author{Charles Wade}
\email{charles.wade@colorado.edu}
\orcid{0000-0002-6056-7717}
\affiliation{%
  \institution{University of Colorado Boulder}
  \city{Boulder}
  \state{Colorado}
  \country{USA}
}
\affiliation{%
  \institution{Draper Scholars, The Charles Stark Draper Laboratory, Inc}
  \city{Cambridge}
  \state{Massachusetts}
  \country{USA}
}

\author{Devon Beck}
\email{dbeck@draper.com}
\orcid{0000-0001-5010-579X}
\affiliation{%
  \institution{The Charles Stark Draper Laboratory, Inc}
  \city{Cambridge}
  \state{Massachusetts}
  \country{USA}
}

\author{Robert MacCurdy}
\email{maccurdy@colorado.edu}
\orcid{0000-0002-1726-151X}
\affiliation{%
  \institution{University of Colorado Boulder}
  \city{Boulder}
  \state{Colorado}
  \country{USA}
}
\authornote{Corresponding Author}

\renewcommand{\shortauthors}{Wade et al.}

\begin{abstract}
Functional gradients give designers finer control over part behavior by varying structure, material, or process conditions across an object. Yet functionally graded fabrication is often treated as a problem of grading geometry or material distribution, rather than grading the slicing and fabrication process itself. In material-extrusion printing, many functional effects arise directly from slicer-controlled mechanisms: local toolpath planning, surface treatment, material assignment, color mixing, and printer state can all vary across space. Mainstream FFF slicers already expose these mechanisms as settings, but they require users to reconstruct heterogeneous intent as manually assigned mesh regions in their GUI. This paper presents slicer project compilation: a fully automated workflow that lowers heterogeneous implicit designs into slicer-native \texttt{.3mf} projects with embedded sub-meshes, settings, recipes, and process-state assignments. The compiler partitions spatial attributes into finite regions, extracts aligned sub-meshes, and serializes those regions into the project dialect required by the target slicer while preserving native toolpath planning, preview, support generation, and printer-profile infrastructure. Our method connects heterogeneous design representations to fully featured slicers, enabling automated and scalable fabrication workflows that retain existing slicer ecosystems. We demonstrate this approach across three classes of slicing and fabrication parameters: settings meshes, virtual extrusion, and color or material halftoning. We also introduce calibrated translation models for temperature responsive foaming TPU and PLA, allowing high-level density and Shore-hardness fields to drive fabrication-ready process fields. Across printed examples, the compiler generates ready-to-slice projects for graded toolpath settings, foaming-filament properties, combined texture and process-state control, and color or material-mixture halftoning, replacing more than 2,500 repetitive manual slicer interactions. Our open-source implementation provides a reusable foundation for functionally graded FFF research and applications built on existing slicer ecosystems.
\end{abstract}

\ccsdesc[500]{Computing methodologies~Volumetric models}
\ccsdesc[100]{Computing methodologies~Mesh geometry models}
\ccsdesc[500]{Applied computing~Computer-aided manufacturing}
\ccsdesc[500]{Applied computing~Computer-aided design}

\keywords{computer-aided design, additive manufacturing, volumetric design, functionally graded materials, slicing}

\begin{teaserfigure}
  \includegraphics[width=\textwidth]{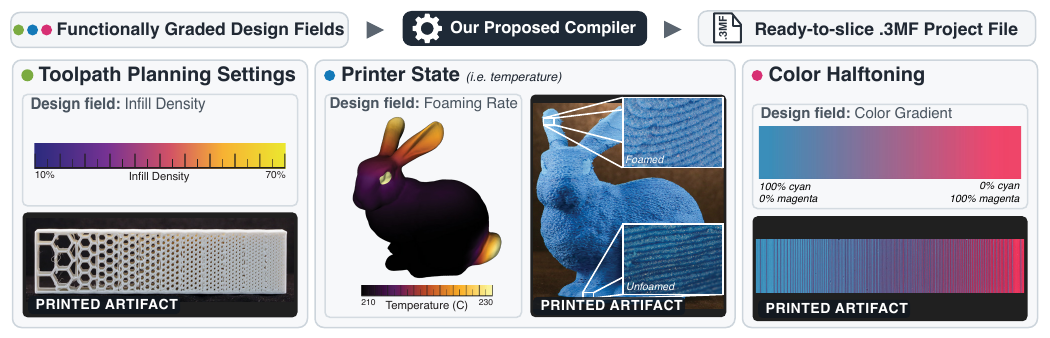}
  \label{fig:teaser}
\end{teaserfigure}


\maketitle

\section{Introduction}
Heterogeneous fabrication aims to make spatial variation a first-class design variable. A printed object may vary stiffness, density, color, surface texture, or infill structure across its volume, and these variations can produce functional behavior that uniform material assignment cannot. Modern additive manufacturing workflows provide many of the low-level mechanisms needed to realize such variation: slicers can assign different settings to different model regions, multi-tool systems can assign different materials, color-mixing workflows can approximate continuous color, and G-code can control printer state during fabrication. These fabrication mechanisms provide the necessary building blocks, but current slicing workflows expose them through interfaces that make dense heterogeneous control impractical: users must manually decompose the design, assign settings region by region, or rely on brittle G-code post-processing. This creates a gap between heterogeneous design methods, which can express continuous spatial intent, and mature slicing tools, which can fabricate local variation only after that intent has been manually entered into the slicer.

Conventional slicers were built around discrete meshes and user-authored settings, not continuous heterogeneous designs. A designer who specifies a volumetric gradient must usually translate that intent into slicer-readable form by hand. In practice, this means partitioning geometry into regions, exporting and aligning many sub-meshes, assigning per-region settings or materials in the slicer interface, and encoding process-state changes through printer-specific scripts or post-processing. This manual lowering ties the design to a particular fabrication setup and limits scalability as the number of regions, attributes, or target slicers increases. Mature slicers provide valuable path planning, preview, support generation, and printer-profile infrastructure, but their interfaces do not directly consume field-based heterogeneous designs.

Heterogeneous modeling is a well-studied upstream design problem, and state-of-the-art field-based representations can encode geometry and spatial attributes as functions over the object domain. Such representations allow a design to express high-level intent, such as target density, Shore hardness, color, infill density, or surface texture, without discretizing that intent into slicer objects. However, the field must still be lowered into the concrete project structures that the target slicer can consume: sub-meshes, setting overrides, material assignments, virtual recipes, logical tools, and printer-specific commands.

To address this gap, we introduce slicer project compilation as an automated lowering step from heterogeneous attribute fields to slicer-native project files. The compiler takes an attribute-modeled implicit design as input, partitions the required spatial fields into finite regions, extracts those regions as aligned sub-meshes, and writes a configured \texttt{.3mf} project with embedded settings, material assignments, recipe definitions, and virtual-tool mappings. The downstream slicer remains responsible for toolpath generation. Thus, the proposed workflow does not replace mature slicers; it leverages them as compilation targets.

The same project-generation framework can target several forms of heterogeneous fabrication control. In this paper, we demonstrate the approach with three representative slicer mechanisms. First, the compiler converts toolpath-planning fields, such as infill density or fuzzy-skin parameters, into region-specific slicer setting overrides. Second, it converts process-state fields into logical tool assignments and custom toolchange commands, allowing slicer-generated toolchange events to control G-code printer states such as nozzle temperature. Third, it converts material-fraction or sRGB color fields into mixed-filament recipe assignments for existing color-mixing workflows. These targets differ in the project data they serialize, but they share the same core abstraction: continuous spatial attributes are reduced to labeled project regions that a conventional slicer can load, preview, and slice.

We evaluate our proposed method with a suite of printed examples that exercise heterogeneous control across slicer-planning settings, process-state assignments, and color or material-mixture workflows. These examples test whether the compiler can preserve spatial fields as slicer-readable project assignments rather than requiring users to recreate those assignments manually in the slicer interface. We also evaluate calibrated translation models used before compilation to resolve high-level property fields into the process controls required for fabrication, allowing object intent to drive machine-state assignments automatically. Across the results, settings-mesh examples demonstrate local control over slicer-generated toolpaths, virtual-extrusion examples demonstrate property-driven process-state control, and halftoning project examples demonstrate compilation of material-fraction and color fields to existing slicer workflows. Together, these results show that slicer project compilation can bridge field-based heterogeneous design and mature slicer workflows without requiring a custom-built slicer for each fabrication task.

\section{Related Work}
\label{sec:related_work}
Prior literature reviews show that heterogeneous design, including volumetric variation and functional grading, can improve printed artifacts and expand the design space of additive manufacturing~\cite{loh_overview_2018,li_review_2020}. A functionally graded object may vary material composition, color, density, stiffness, surface texture, toolpath structure, or process state across space. This breadth creates a translation problem: heterogeneous intent can originate in field-based design representations, but fabrication may require local composition control, slicer setting overrides, printer-state control, or color halftoning. We therefore review prior work in four areas that define this interface: field-based representations and compilation, local composition control, spatial control of slicer settings and process state, and color halftoning for material extrusion. Across these areas, we focus on the gap our work addresses: using mature slicers and readily available desktop FFF printers to realize heterogeneous designs without custom slicers, manually assigned regions, or post-processed G-code.

\subsection{Field-Based Design Representations and Compilation}
\label{sec:rw_field_based_design_compilation}

Field-based design representations support heterogeneous design by specifying spatial information that boundary meshes represent poorly. A mesh can describe an object's exterior surface, and multiple meshes can describe discrete regions, but a smooth volumetric gradient must be approximated by many separate boundaries before a conventional slicer can assign materials or settings. Field-based representations avoid this reconstruction step by describing geometry, composition, color, or other attributes as functions over space. Programmable fabrication systems such as OpenFab use this idea to express material variation procedurally~\cite{vidimce_openfab_2013}, while OpenVCAD represents implicit geometry and volumetric material composition as spatial fields for multi-material fabrication~\cite{wade_openvcad_2024,wade_implicit_2025}. These representations let designers describe heterogeneous intent directly, but fabrication still requires a compiler that converts fields into the concrete inputs expected by a target manufacturing workflow.

Recent heterogeneous attribute modeling extends this field-based view by treating heterogeneous fabrication as a multi-stage compilation problem~\cite{wade_attribute_modeling}. Rather than representing only geometry or material composition, the approach describes objects using named, typed spatial attributes that may encode high-level design intent, such as target mechanical response, hardness, or color. Translation models then map these intent-level attributes into process-specific realization fields, such as material recipes. This framing is important for our work because it separates what the designer specifies from the machine-specific structures that fabrication requires. The remaining compilation step must convert the resolved realization fields into the concrete project data consumed by the target workflow.

\subsection{Local Composition Control for Functionally Graded Fabrication}
\label{sec:rw_local_composition_control}

Local composition control provides the most direct route to functionally graded fabrication. In these systems, the printer changes the relative amounts of two or more base materials across an object, producing spatial variation in appearance or material behavior. Voxel-based material jetting demonstrates this idea at high spatial resolution by assigning material channels throughout a volume, and programmable fabrication pipelines have used this capability to express complex multi-material distributions~\cite{vidimce_openfab_2013,wade_openvcad_2024}. These systems establish the value of volumetric material control but focus only on material-jetting workflows and are not directly applicable to FFF slicing control.

Material-extrusion systems approach local composition control through a different mechanism. Multi-feed and mixing hotend systems combine filaments during deposition, allowing the commanded mixture ratio to vary during a print. Kennedy and Christ demonstrate that active in-situ mixing can blend polymer filaments during fused filament fabrication~\cite{kennedy_printing_2020}. Similarly, Green et al. use an active-mixing hotend to spatially control mixture ratios and fabricate graded compositions~\cite{green_local_2023}. These works establish mixing extrusion as a viable route for graded composition in FFF. They also motivate the fabrication-planning problem that follows: a spatial composition field must still be converted into printer instructions that coordinate material ratios with the toolpath. Garland and Fadel demonstrate this challenge on an off-the-shelf mixing extrusion system by discretizing a gradient into regions, assigning those regions as virtual extruders in Slic3r, and post-processing the resulting G-code to replace toolchange commands with mixture-ratio commands~\cite{garland_design_2015}. Leoni et al. further frame this problem as a gap between functionally graded design and material-extrusion fabrication~\cite{leoni_functionally_2023}. These works highlight that local composition control in FFF depends not only on extrusion hardware, but also on representations and software workflows that can translate graded design intent into slicer-compatible assignments.

Direct-ink-write systems provide another important class of local composition control. Active mixing has enabled graded fluids, reactive materials, ceramics, elastomers, and dielectric composites~\cite{ober_active_2015,ortega_active_2019,pelz_multi-material_2021}. Duncan et al., for example, use active mixing of nanocomposite inks to fabricate low-loss graded dielectrics with spatially tailored electromagnetic properties~\cite{duncan_lowloss_2023}. These examples show that continuous material variation can produce application-specific functional behavior. However, they also show that DIW composition grading is typically developed as an application-specific fabrication workflow, where the material formulation, printhead behavior, path design, and process commands must be coordinated for each system rather than specified through a reusable heterogeneous design.

These systems establish local composition control as a useful fabrication primitive. The remaining challenge is how to translate a spatial composition field into slicer-compatible regions and material assignments without requiring the designer to reconstruct the gradient as a collection of manually assigned meshes when moving between tools.

\subsection{Spatial Control of Slicer Settings and Process State}
\label{sec:rw_slicer_settings_process_state}

In addition to composition variation, slicers also shape local behavior through the toolpaths they generate: infill density, perimeter count, surface texture, and related settings can change stiffness, strength, weight, and appearance while using a single material. Prior work shows that toolpath structure and infill design strongly affect mechanical performance, and several methods use stress, topology, or other fields to guide path orientation and material placement~\cite{xia_stress-based_2020,sales_function-aware_2021,liu_stress_2024}. These methods make toolpath planning itself part of the design space. Modern slicers expose a related capability through object-level settings, modifier meshes, and painted regions, which allow users to apply different path-generation rules to different parts of a build~\cite{prusa_per_model_settings}. However, these interfaces remain largely interactive. A designer can assign different infill densities to different imported meshes, but a high-resolution spatial field must first be reconstructed as many aligned models and then assigned settings region by region.

Surface texture provides a compact example of the same issue. Fuzzy skin perturbs exterior toolpaths to produce a roughened surface, and slicers expose this behavior as a global setting or as a locally painted surface effect~\cite{prusa_fuzzy_skin, vesco_fuzzy_2025}. Thus, surface texture is a slicer-level property rather than a material-composition property. Yet the slicer interface does not provide a general field representation for smoothly varying texture parameters over a model. As with infill, the available slicing feature is useful for functional grading, but the interface between spatial design intent and path-generation parameters remains indirect.

Process-state control broadens this pattern beyond slicer settings. A printer can vary local properties by changing machine commands during fabrication, including flow rate, print speed, cooling, laser power, energy input, or nozzle temperature. Metal additive manufacturing has used site-specific process control to vary melt-pool behavior, geometry, porosity, and material response~\cite{gibson_beyond_2019,borish_automated_2022,yuan_effect_2022}. In material extrusion, foaming filaments provide a single-material route to graded density because extrusion temperature and flow compensation affect expansion, porosity, and mechanical response~\cite{damanpack_porous_2021,tammaro_microfoamed_2022,ozdemir_xpandables_2023}. These examples show that functional grading can come from printer state, in addition to material mixtures and toolpath settings.

However, process-state fields introduce a planning problem that ordinary slicer controls rarely address. Some commands can change instantaneously, while others require stabilization time, material flow, or thermal response before the printed material reaches the intended state. Gradient-informed slicing addresses this problem by making the spatial field part of toolpath planning, including for mixture ratios, tool changes, and temperature responsive foaming materials~\cite{wade_implicit_2025-1}. This approach shows the value of custom gradient-aware slicers, but it also illustrates a tradeoff: a custom slicer must reimplement many surrounding capabilities that mature slicers already provide, including support generation, skin and infill strategies, printer profiles, modifier semantics, and stable project workflows.

Our proposed method follows the complementary direction of working within full-featured slicer workflows rather than replacing them. Existing slicers already contain abstractions for local control, including object-level settings, modifier regions, tool assignments, and tool-change events. Prior gradient-informed workflows also show that tool-like channels can represent fabrication states rather than only physical extruders~\cite{wade_implicit_2025-1}. The open problem is how to connect spatial design fields to these slicer abstractions automatically, without reducing the workflow to manual region assignment or requiring a bespoke slicer for each graded fabrication task.

\subsection{Color Halftoning for Material Extrusion}
\label{sec:rw_color_halftoning}

Color provides another instance of spatially varying fabrication intent, but it differs from the process-state and slicer-setting controls because the color reproduction problem has a long computational history. Full-color 3D printing must map continuous colors to a finite set of printable materials or deposition states. Prior work has addressed this mapping through color management, contoning, surface halftoning, voxel assignment, and line-based halftoning~\cite{yuan_accurate_2021,babaei_color_2017}. In material extrusion, Reiner et al. approximate continuous tones with dual-color deposition patterns~\cite{reiner_dualcolor_2014}, while Kuipers et al. adapt hatching and line-based halftoning to the path structure of fused deposition modeling~\cite{kuipers_3d_2017,kuipers_hatching_2018}. Song et al. further show colored fused filament fabrication using layered color strata and mixing-nozzle hardware~\cite{song_colored_2019}. These methods show that discrete filaments or material states can approximate continuous color, but they focus primarily on the color assignment, optical model, or deposition strategy.

Recent slicers make filament-based color workflows more accessible by exposing color-mixing and halftoning features directly in open-source slicing ecosystems. Prusa ColorMix and Orca FullSpectrum provide practical mechanisms for assigning printable color mixtures within slicer workflows~\cite{prusa_colormix_2026,ratdoux_orcaslicer_fullspectrum_2026}. These systems are important because they turn color halftoning from a standalone research pipeline into a slicer-level fabrication capability. However, the remaining interface problem mirrors the challenges with local settings and process state: a full-color design must still be converted into slicer-level assignments that the downstream halftoning method can consume.

This places color within the same interface gap as other heterogeneous fabrication parameters. Prior work provides effective color-assignment and halftoning methods, and recent slicers provide practical mechanisms for fabricating those assignments. The missing step is a general way to connect full-color heterogeneous designs to slicer-level color workflows without requiring the designer to manually reconstruct the design as printable regions or recipes.

\subsection{Summary}
The challenge with heterogeneous fabrication workflows is neither field representations nor slicer features in isolation. Prior work provides rich design mechanisms for local composition control, process-state control, color reproduction, and heterogeneous modeling. Existing slicers also provide mature path-generation features and local control mechanisms. The remaining challenge is an automated lowering step that connects field-based heterogeneous intent to a format a mature slicing program can ingest. This work addresses that gap by generating slicer-ready projects in which the sub-meshes, settings, virtual tool assignments, and color assignments are embedded before slicing.

\section{Methods}
\label{sec:methods}

\subsection{Overview: Slicer Project Compilation}
\label{subsec:methods_overview}

We define slicer project compilation as the conversion of an attribute-modeled implicit design into a slicer-native \texttt{.3mf} project whose sub-regions carry the settings, process-state assignments, or material-mixture recipes required by the target workflow. The slicer project compiler does not replace the downstream slicer. Instead, it outputs a project file that the slicer can load, preview, slice, and convert to G-code using its native toolpath generator.

Our proposed compiler takes as input an implicit heterogeneous object whose compiler-required attributes have been authored directly or resolved before compilation. Required attributes are the fields that a selected compilation target needs in order to generate slicer-ready project data, and they vary by compilation type. For example, one compiler may require slicer-planning settings such as infill density, another may require process-state setpoints such as nozzle temperature or flow-rate compensation, and another may require color or material-fraction fields. The compiler first partitions one or more attributes into a finite set of spatial regions. It then extracts each nonempty region as a sub-mesh and labels that sub-mesh with representative attribute values. The final serialization step is target-specific: the compiler writes the metadata, tool assignments, recipe definitions, and project settings required by the selected slicer-project dialect.

\begin{figure*}
    \centering
    \includegraphics[width=0.8\linewidth]{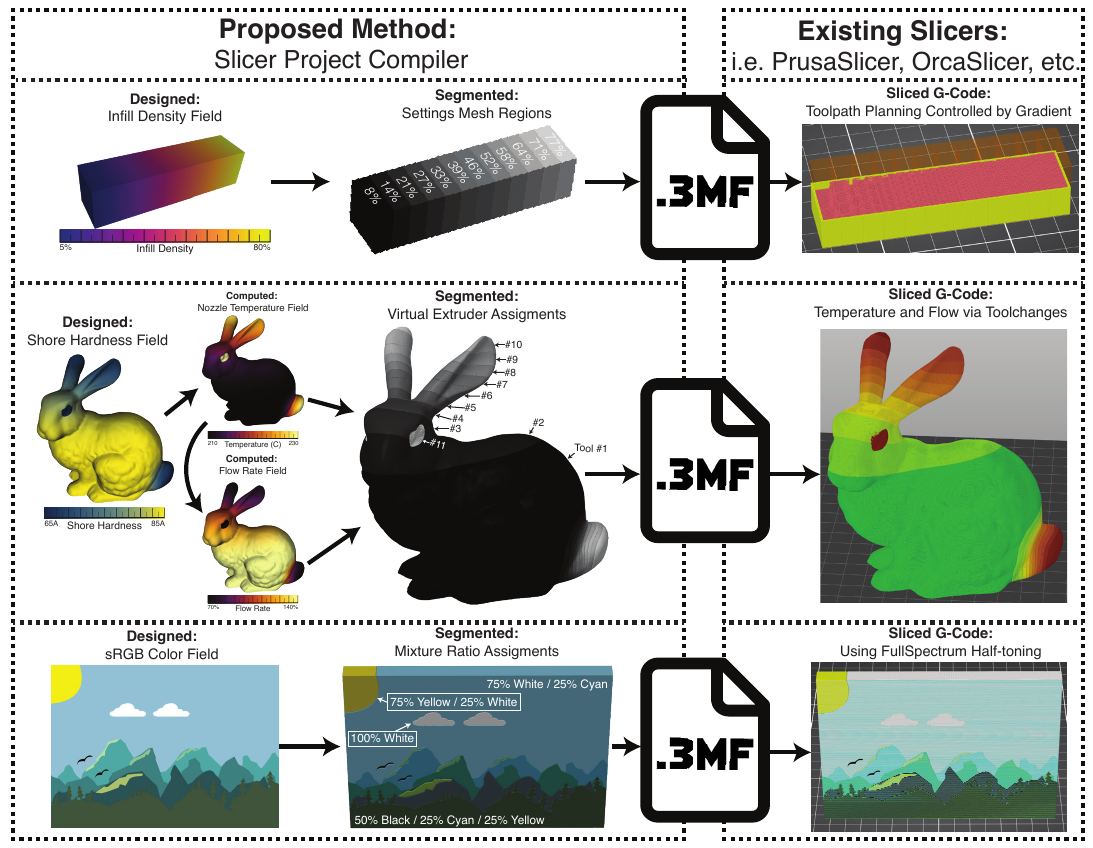}
    \caption{Slicer project compilation lowers heterogeneous attribute fields into slicer-native \texttt{.3mf} projects. The compiler partitions resolved fields into labeled sub-meshes and serializes them as settings-mesh regions, virtual-extruder assignments, or mixture-ratio assignments. Existing slicers then perform their native toolpath planning, toolchange emission, or halftoning workflows from the generated project files.}
    \label{fig:overview}
\end{figure*}

Our compiler abstraction is agnostic to the selected compilation target. As shown in Figure~\ref{fig:overview}, we demonstrate this abstraction with three compiler targets, while preserving a shared partitioning and sub-mesh generation stage. Settings meshes assign slicer-planning parameters, such as infill density or fuzzy-skin settings, to generated regions before toolpath planning. Virtual extrusion assigns regions to logical tools and uses custom toolchange templates to redirect slicer toolchange events into process-state commands. Color and material halftoning projects assign regions to virtual mixed-filament recipes that ColorMix or FullSpectrum-style workflows convert into filament-level halftoning. These targets differ in the assignments they write, but they share the same field partitioning and sub-mesh generation stage.

The targets can also be combined when the slicer-project dialect supports the required assignments. For example, a foaming-filament workflow may use virtual extrusion to control temperature and flow rate, while also using settings meshes to apply spatially varying fuzzy skin. In such cases, the compiler partitions the participating attributes jointly so that each generated region carries an identical assignment tuple.




\subsection{Compiler Input and Attribute Resolution}
\label{subsec:input_representation}

The slicer project compiler consumes heterogeneous implicit designs produced by an upstream attribute-modeling workflow~\cite{wade_attribute_modeling}. Each input design provides an implicit solid geometry together with named spatial attribute fields defined over the object. These attributes may describe slicer-planning settings, process-state setpoints, color, material fractions, or higher-level design intent.

The selected slicer-project target determines which spatial fields must be available before project generation. Settings-mesh targets consume slicer-planning attributes such as infill density or fuzzy-skin parameters. Virtual-extrusion targets consume process-state attributes such as nozzle temperature and flow-rate compensation. Color and material halftoning targets consume color fields or material-fraction fields. Our proposed compiler then lowers these resolved fields into slicer-native project assignments.

Consistent with heterogeneous implicit design methods~\cite{wade_attribute_modeling}, required process fields may be authored directly or be resolved from higher-level intent before compilation. For example, a design may specify target density or Shore hardness rather than nozzle temperature. In addition to our slicer compiler, we introduce new translation models to resolve printer process parameters from design intent. As we show in section \ref{subsubsec:results_foaming_calibration}, our calibrated translation models can compute the temperature and flowrate fields for fabrication with foaming filaments.

This separation between translation models and slicer-project compilation keeps the compiler target-specific but design-source agnostic. The same compiler can consume directly authored slicer settings, translated process fields, or color and material-fraction fields, provided that the selected target receives the attributes it requires. The remaining methods therefore focus on how those resolved fields are partitioned, converted into sub-mesh regions, and serialized into slicer-native \texttt{.3mf} projects.

\subsection{Slicer-Native \texorpdfstring{\texttt{.3mf}}{3MF} Projects as Compiler Outputs}
\label{subsec:3mf_projects}

A slicer-native \texttt{.3mf} project records the fabrication setup that a slicer normally builds through interactive use. In a conventional workflow, a designer imports one or more meshes, transforms and arranges them on the build plate, selects global print settings, applies model- or region-level overrides, and assigns materials or extruders. The slicer can then save this configured job as a single project file for later editing, sharing, or slicing. The project therefore contains more than boundary geometry: it stores the relationship between geometric objects and the slicer state needed to fabricate them.

This makes a slicer-native \texttt{.3mf} project a useful compiler target. A conventional mesh file can describe surfaces, but it cannot express the slicer settings, tool assignments, or mixed-filament recipes required by a heterogeneous design. A project file can carry this missing context in a form the slicer interprets. The slicer project compiler therefore automates the interactive setup process: it generates the required sub-meshes, attaches the appropriate per-region assignments, writes the project-level configuration, and emits a configured fabrication job rather than an unannotated collection of meshes.

Different slicers serialize these project components using different dialects. PrusaSlicer-style projects and Orca/Bambu component-style projects both store geometry, project settings, and per-region assignments, but they organize those data differently and use different metadata conventions. The slicer project compiler separates the abstract assignment produced by the method from the dialect-specific serialization written to disk.

\subsection{Attribute Partitioning}
\label{subsec:attribute_partitioning}

\begin{figure}
    \centering
    \includegraphics[width=1.0\linewidth]{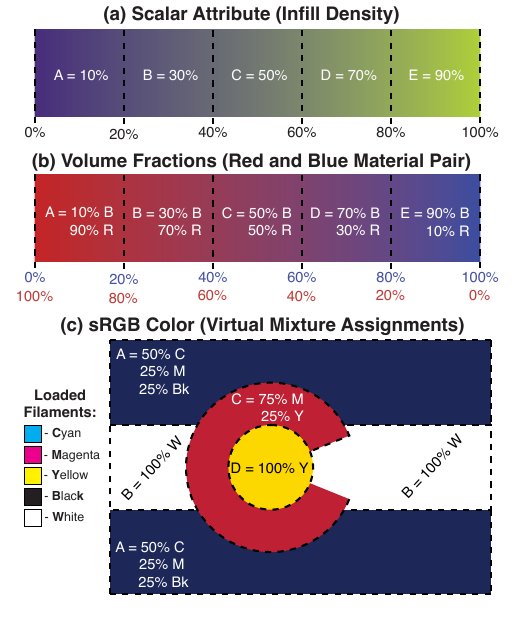}
    \caption{Attribute partitioning converts continuous heterogeneous fields into finite slicer-project assignments. (a) Scalar fields are divided into representative intervals, shown here for infill density. (b) Material-fraction fields are partitioned in recipe space so each region carries a valid mixture vector. (c) Color fields are mapped to selected virtual mixture assignments derived from the loaded filament set.}
    \label{fig:partitioning}
\end{figure}

As detailed in Figure~\ref{fig:partitioning}, all three slicer project targets use the same geometric reduction: the compiler converts continuous attribute fields over the object interior into a finite set of labeled sub-meshes. The compiler samples the implicit geometry and the required attributes on a regular grid and assigns sampled points to discrete attribute regions, where a region represents a range of attribute values. The compiler then constructs a region-indicator mask for each region, intersects that mask with the object domain, and extracts the resulting boundary with marching cubes~\cite{lorensen_marching_1987}. The output is a collection of sub-meshes whose labels form a piecewise-constant approximation of the resolved functionally graded attributes. Each sub-mesh is assigned a single attribute value.

The compiler extracts each region as a sub-object. For a selected attribute bin or recipe, it defines a region-membership field whose zero set marks the boundary between points assigned to that region and points assigned elsewhere. The compiler combines this field with the object SDF so that the extracted surface is clipped to the original solid domain. Marching cubes then extracts the boundary of this sub-object as a triangle surface mesh. As a result, the region meshes can place boundaries between grid samples when the geometry and attribute fields vary smoothly, even though the compiler samples the fields on a regular grid. The chosen region count sets the fidelity--cost tradeoff: more regions better preserve spatial variation, but produce more sub-meshes, larger project files, and more downstream slicer work.

\subsubsection{Scalar Attribute Partitioning}
\label{subsubsec:scalar_partitioning}

Scalar partitioning supports settings meshes, virtual extrusion, and any workflow that must assign one or more scalar controls jointly. Given scalar fields, the compiler first measures each field's observed range over sampled points inside the object. It then divides each range into $N$ equal-width intervals and uses each interval midpoint as the representative value assigned to regions in that interval. 

When several scalar attributes must vary together, the compiler constructs the Cartesian product of their intervals. A point belongs to a product cell only if every participating attribute falls within the corresponding interval. This joint partition ensures that each generated sub-mesh carries an identical tuple of representative values, such as a temperature and flow-rate pair or a fuzzy-skin setting together with a virtual-extrusion state. With $m$ attributes and $N$ intervals per attribute, the method can produce up to $N^m$ candidate regions. 

Algorithm~\ref{alg:scalar_attribute_partitioning} summarizes this process. For settings meshes, the representative labels become slicer-setting values. For virtual extrusion, they become logical process-state assignments. For combined workflows, the labels may include both planning settings and process-state values.

\begin{algorithm}[th]
\caption{Partition an Implicit Object by Scalar Attribute Fields}
\label{alg:scalar_attribute_partitioning}
\KwIn{Implicit geometry; scalar attribute fields; intervals $N$ per attribute}
\KwOut{Labeled sub-mesh regions with representative scalar values}
\BlankLine

Sample the geometry and scalar attributes inside the object\;

\ForEach{attribute field $A_i$}{
    Compute its sampled value range\;
    Partition the range into $N$ equal-width intervals\;
    Store each interval midpoint as its representative value\;
}

Form the Cartesian product of intervals across all attributes\;

\ForEach{interval tuple $B$}{
    Select the object region whose attribute values fall in $B$\;
    Intersect the selected region with the implicit geometry\;
    Extract its boundary as a sub-mesh\;

    \If{the sub-mesh is nonempty}{
        Label it with the representative values from $B$\;
        Add it to the output\;
    }
}

Return the labeled sub-meshes\;
\end{algorithm}

\subsubsection{Material-Fraction Partitioning}
\label{subsubsec:volume_fraction_partitioning}

Material-fraction fields require a different value-space partition because each sample is a recipe vector rather than an independent scalar. For $k$ materials, each sampled material-fraction value is a nonnegative
length-\(k\) vector whose components sum to one. Component-wise scalar binning would create many invalid or redundant recipe combinations, so the compiler partitions the sampled vectors directly in material-fraction space.

The compiler divides material-fraction space into a bounded set of representative recipes. Unlike an arbitrary scalar field, a material-fraction field has known bounds: each component lies between 0 and 1, and the components form a normalized fraction vector. The compiler can therefore partition the full feasible recipe space directly. For a two-material field ($k=2$) with materials A and B, choosing $N=4$ creates four fraction ranges: 0--25\% A with 100--75\% B, 25--50\% A with 75--50\% B, 50--75\% A with 50--25\% B, and 75--100\% A with 25--0\% B. The compiler assigns each sampled material-fraction vector to the nearest range, computes a representative fraction vector for each occupied range, and maps that vector to a printable recipe. The resulting spatial regions are then extracted with the same geometry-intersection and marching-cubes procedure used for scalar fields.

Algorithm~\ref{alg:volume_fraction_partitioning} describes this procedure. The key distinction from scalar partitioning is that the partition is built in the attribute's natural value space: scalar intervals for scalar fields, and median-cut recipe regions for material-fraction vectors.

\begin{algorithm}[th]
\caption{Extend Attribute Partitioning to Material Volume Fractions}
\label{alg:volume_fraction_partitioning}
\KwIn{Implicit geometry; material-fraction field; maximum regions $N$}
\KwOut{Labeled sub-mesh regions with representative material recipes}
\BlankLine

Sample the material-fraction field inside the object and normalize each vector\;

Partition material-fraction space into $N$ groups\;

\ForEach{group $G_j$}{
    Compute a representative fraction vector $\bar{v}_j$\;
}

Assign each object point to the nearest representative $\bar{v}_j$\;

\ForEach{representative $\bar{v}_j$}{
    Form the spatial region assigned to $\bar{v}_j$\;
    Extract its nonempty sub-mesh using Algorithm~\ref{alg:scalar_attribute_partitioning}\;
    Label the sub-mesh with material recipe $\bar{v}_j$\;
}

Return the labeled material-recipe sub-meshes\;
\end{algorithm}

\subsubsection{Color-Field Partitioning and Palette Selection}
\label{subsubsec:color_partitioning}

Color fields require an additional selection step because the loaded filaments cannot reproduce every sRGB value equally well. If the compiler assigned each sampled color independently, it could request many recipes whose predicted colors are poor approximations or whose small differences provide little visible benefit. We therefore select a bounded printable palette before spatial region extraction. The palette contains recipes that cover the colors present in the object, given the fixed physical filaments and the target slicer's color-mixing model.

\begin{figure}
    \centering
    \includegraphics[width=1.0\linewidth]{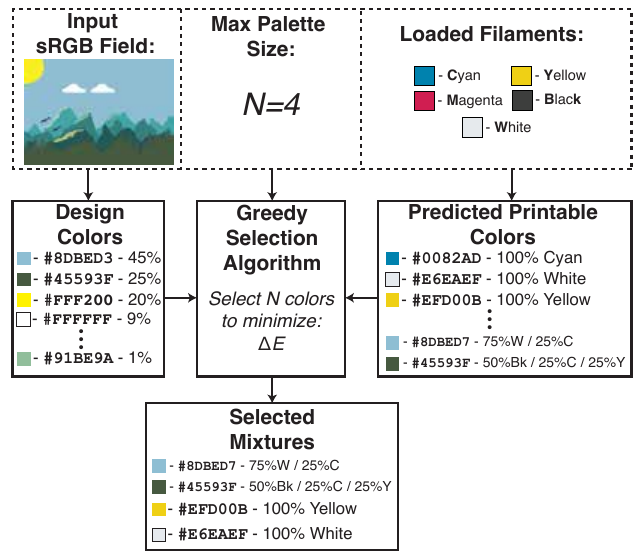}
    \caption{Palette selection for compiling an sRGB color field into printable mixture assignments. The compiler samples the design colors, predicts the apparent colors of candidate recipes from the loaded filaments, and greedily selects at most \(N\) mixtures that minimize weighted perceptual color error. The selected mixtures become the labels used for downstream color-region extraction and project serialization.}
    \label{fig:color_matching}
\end{figure}

As shown in Figure~\ref{fig:color_matching}, the input is an sRGB color field and a fixed set of loaded physical filaments with specified RGB colors. The compiler internally generates a discrete vocabulary of printable recipes containing up to three filaments at fixed proportions. The downstream slicers limit recipes to three filaments. Workflow-specific color-prediction models estimate the apparent color of each candidate recipe. In this paper, the ColorMix target uses \texttt{prusa-fdm-mixer}, while the FullSpectrum target uses \texttt{filament-mixer}. These models predict recipe appearance; they do not generate the recipe vocabulary.

The compiler aggregates sampled target colors into a weighted distribution and greedily selects at most $N$ candidate recipes that minimize sample-count-weighted CIEDE2000 error, $\Delta E_{00}$. Each sampled point is assigned to the selected recipe with the smallest predicted $\Delta E_{00}$ from the requested color $C(p)$. These assignments define color-recipe regions, which are extracted as sub-meshes using the same geometric procedure as the scalar and material-fraction cases.

Algorithm~\ref{alg:color_slicer_project} summarizes full-color slicer project compilation. The algorithm differs from the previous two only in how it chooses the representative labels. Once the selected printable recipes exist, the spatial extraction step again produces labeled sub-meshes that can be serialized into the target slicer-project dialect.

\begin{algorithm}
\caption{Compile a Full-Color Attribute Field into a Slicer Project}
\label{alg:color_slicer_project}
\KwIn{Implicit geometry; sRGB color field; filament palette;
      candidate recipes; target workflow; maximum palette size $N$}
\KwOut{Color-assigned sub-meshes in a 3MF project and a color report}
\BlankLine

Sample $C$ inside the object and aggregate similar colors into a frequency distribution\;

\ForEach{candidate recipe}{
    Predict its printable color\;
}

Greedily select at most $N$ recipes that minimize frequency-weighted perceptual color error\;

\ForEach{sampled point $p$ inside the object}{
    Assign its color to the selected recipe with the closest predicted color\;
}

\ForEach{selected recipe}{
    Form the spatial region assigned to that recipe\;
    Extract its nonempty sub-mesh using Algorithm~\ref{alg:scalar_attribute_partitioning}\;
    Label the sub-mesh with the corresponding filament or recipe\;
}

Package filament definitions, printable mixtures, color-assigned sub-meshes,
and slicer metadata into a 3MF project\;

Return the 3MF project and color report\;
\end{algorithm}

\subsection{Slicer-Planning Control with Settings Meshes}
\label{subsec:settings_meshes}
\begin{figure*}[t]
    \centering
    \includegraphics[width=\textwidth]{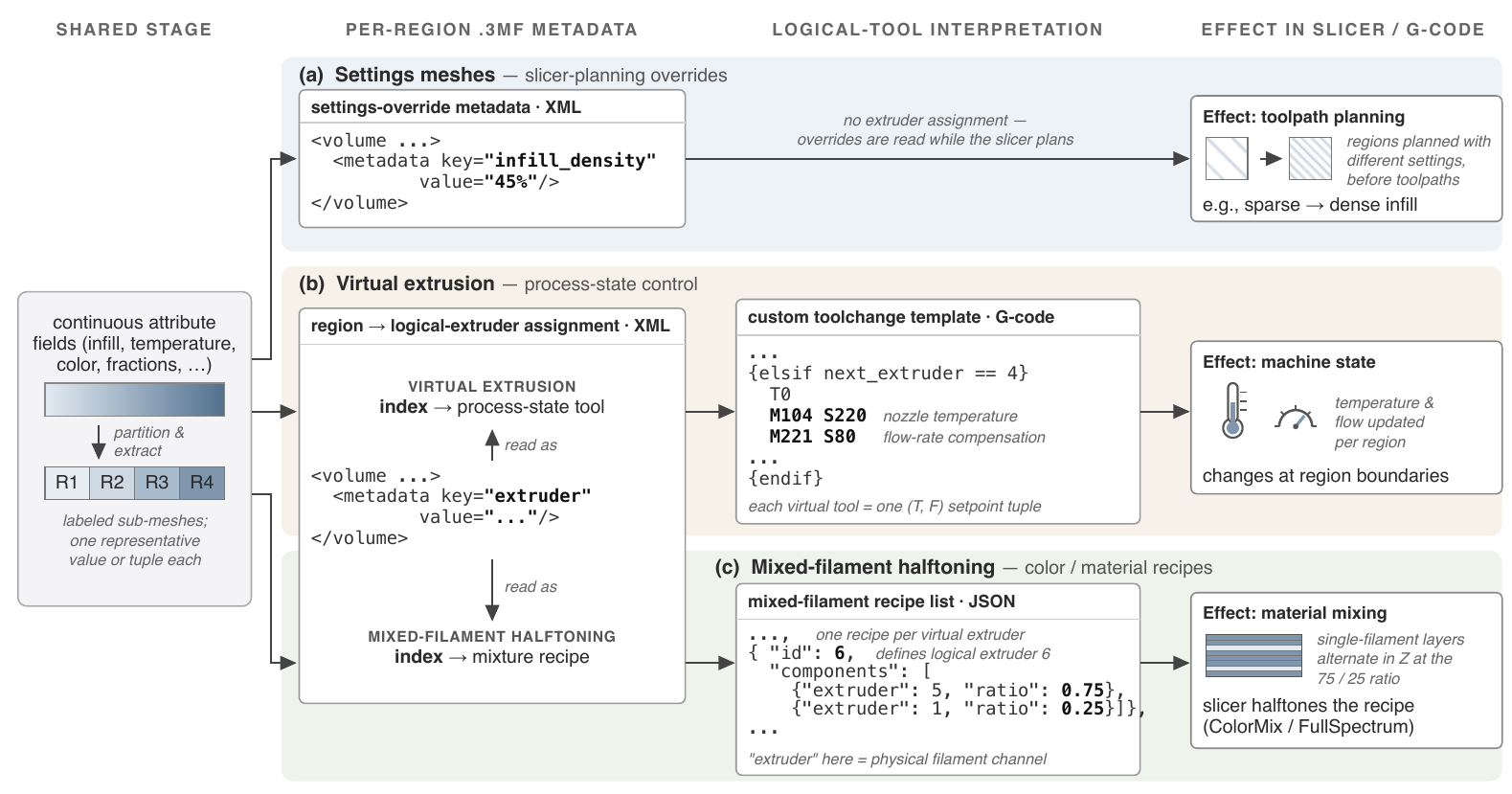}
    \caption{The three slicer-project compilation targets write different
    per-region assignments from the same partitioned sub-meshes. Settings
    meshes (a) attach slicer-planning overrides directly to region metadata
    and involve no tool assignment. Virtual extrusion (b) and mixed-filament
    halftoning (c) share the region-to-logical-extruder assignment but
    interpret the extruder index differently: (b) resolves it through a custom
    toolchange template into process-state G-code such as \texttt{M104}
    (temperature) and \texttt{M221} (flow rate), while (c) resolves it to an
    entry in a recipe list that the slicer halftones from the loaded physical
    filaments. Excerpts are abbreviated; the compiler writes the full native
    dialect of the target slicer.}
    \label{fig:slicer_project_mechanisms}
\end{figure*}

Settings meshes provide direct control over slicer-planning parameters by encoding partitioned scalar attributes as slicer-recognized regions with setting overrides. These meshes are generated directly from the partitioning process described in Algorithm~\ref{alg:scalar_attribute_partitioning}. In the emitted \texttt{.3mf} project, each region is represented as XML metadata in the slicer's native dialect. Figure~\ref{fig:slicer_project_mechanisms}a shows a representative example for PrusaSlicer in which a region mesh is assigned an infill-density override. Because these values exist before toolpath generation, they can change the slicer's planning behavior rather than post-process the generated G-code.

Settings meshes only apply to parameters that the slicer already exposes as model- or volume-level overrides. Examples include infill density, fuzzy skin parameters, speeds, extrusion width, and perimeter count. 
This component of our workflow automates a process that would otherwise require manual project construction. Without slicer project compilation, a designer must export separate region meshes, import them into the slicer, align them, and assign each setting override through the user interface. The proposed compiler instead emits one configured \texttt{.3mf} project in which the region geometry and settings assignments are embedded. We report the number of user interactions saved by using our proposed method compared to manual assignment in Section \ref{subsec:results_interaction_savings}.


\subsection{Machine-State Control with Virtual Extrusion}
\label{subsec:virtual_extrusion}

Settings meshes control parameters that the slicer already exposes as per-region planning settings. Virtual extrusion addresses a different case: spatially varying process states that the slicer does not ordinarily expose as local model settings. We represent each discrete process-state tuple as a logical tool. The slicer then emits toolchange events between those logical tools, and a custom toolchange template converts each event into G-code commands that update machine state.

\begin{figure}
    \centering
    \includegraphics[width=1.0\linewidth]{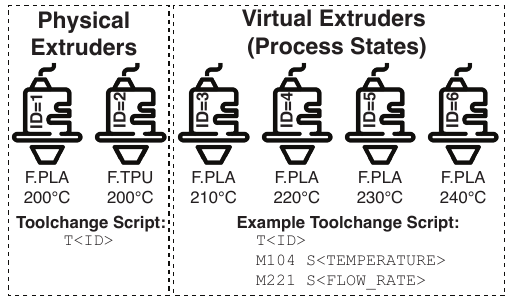}
    \caption{Virtual extrusion maps logical slicer tools to process-state setpoints rather than only physical extruders. Physical extruders correspond to real filament channels, while additional virtual extruders encode states such as nozzle temperature and flow-rate compensation. Custom toolchange scripts (Fig.~\ref{fig:slicer_project_mechanisms}b) convert slicer toolchange events into the G-code commands required to apply each process state.}
    \label{fig:virtual_extrusion}
\end{figure}

As shown in Figure \ref{fig:virtual_extrusion}, virtual extrusion separates physical extruders from virtual extruders. A physical extruder is an actual printer tool, nozzle, or filament channel. A virtual extruder is a logical tool index used by the slicer to identify a process-state setpoint or setpoint tuple. The printer may still have one physical nozzle, but the project can declare additional logical tools whose toolchange events map to commands such as temperature or flow-rate updates. In this paper, we demonstrate this mechanism with temperature and flow rate, using \texttt{M104} to set nozzle temperature and \texttt{M221} to set flow-rate compensation. However, the same mechanism can emit any G-code sequence that the printer accepts.

The compiler applies Algorithm~\ref{alg:scalar_attribute_partitioning} to the process-state fields. Each occupied scalar interval, or joint interval for multiple process-state fields, receives a representative tuple such as $(T,F)$, where $T$ is temperature and $F$ is flow-rate compensation. The compiler assigns each resulting sub-mesh to the logical tool associated with that tuple. The slicer sees these regions as ordinary tool assignments, but the custom toolchange template interprets the next logical tool index as a request to update machine state. Figure~\ref{fig:slicer_project_mechanisms}b shows this composition: each region carries a logical-extruder assignment, and the template converts the resulting toolchange events into the corresponding process-state commands.

This mechanism is useful for foaming filaments because the local expansion state depends on nozzle temperature, while geometric accuracy often requires compensatory flow control. A pre-compilation translation model can therefore resolve a desired density or hardness field into temperature and flow-rate fields, and virtual extrusion can then express those resolved process fields in a slicer project without requiring a custom slicer or custom path generator. Table~\ref{tab:settings_vs_virtual_extrusion} summarizes the distinction between settings meshes and virtual extrusion.

\begin{table}
\centering
\caption{Settings meshes and virtual extrusion both use scalar partitioning,
but they affect different stages of the fabrication workflow.}
\label{tab:settings_vs_virtual_extrusion}
\begin{tabularx}{\columnwidth}{@{}p{0.30\columnwidth}XX@{}}
\toprule
\textbf{Method} & \textbf{Controls} & \textbf{When applied} \\
\midrule
Settings meshes &
Slicer planning parameters, such as infill, speeds, widths, and fuzzy skin. &
Before toolpath generation; the slicer uses the values during planning. \\
Virtual extrusion &
G-code-controllable process states, such as temperature and flow rate. &
At logical toolchange events in the generated G-code. \\
\bottomrule
\end{tabularx}
\end{table}



\paragraph{Ordering virtual tools for slow process states.}
Some virtual extrusion workflows also require an ordering policy. This is not necessary for every G-code-controllable state, but it matters when transitions have time or material costs. Temperature control for foaming filaments is one such case: the nozzle needs time to reach a new temperature, and the extruded material may require purging or a prime tower before the new foaming state stabilizes.

For temperature-based virtual extrusion, the compiler indexes virtual tools in increasing temperature. We then use a reversing layer order: the slicer visits tools in ascending order on one layer and descending order on the next, for example $1,2,3,4,5$ followed by $5,4,3,2,1$. This preserves monotonic local changes while avoiding a large reset from the hottest state to the coolest state at each layer boundary.

\subsection{Filament-Mixture Control with Color and Material Halftoning Projects}
\label{subsec:color_material_halftone_projects}

ColorMix and FullSpectrum-style slicer workflows provide a third form of slicer project control: they approximate intermediate colors or material recipes by assigning regions to virtual mixed-filament recipes. A printer loads a fixed set of physical filaments, such as CMYKW, and the slicer defines virtual recipes that combine one, two, or three of those filaments at specified ratios. The slicer then performs the downstream halftoning or filament-mixing procedure needed to realize those recipes. The slicer project compiler uses this mechanism by generating recipe definitions and assigning extracted sub-meshes to the corresponding virtual recipes.

This workflow supports two input cases. If the resolved design contains a material-fraction field, the compiler uses Algorithm~\ref{alg:volume_fraction_partitioning} directly. It partitions the field into representative material-fraction regions, maps each component of the volume-fraction vector to a loaded filament channel, and writes the project metadata that defines the mixture ratio for each generated sub-mesh. This path bypasses color matching because the design already specifies the desired material mixture rather than an appearance target.

If the resolved design instead contains an RGB color field, the compiler treats the loaded filament set as a constrained color gamut. It does not assume that arbitrary sRGB values can be reproduced accurately. Instead, as described in Algorithm~\ref{alg:color_slicer_project}, the compiler generates a discrete vocabulary of printable recipes, predicts each recipe's apparent color with the workflow-specific color model, and selects a bounded palette that best covers the sampled target colors under a sample-count-weighted $\Delta E_{00}$ objective. It then assigns each spatial region to the selected recipe with the smallest predicted perceptual error.

Prusa ColorMix and Orca FullSpectrum use the same abstract representation of physical filaments, virtual mixtures, and region assignments, but they serialize those data differently and use different downstream half-toning algorithms. The compiler therefore treats them as separate project dialects under one method: both consume labeled sub-meshes and recipe definitions, but each requires its own native metadata. Figure~\ref{fig:slicer_project_mechanisms}c shows a representative entry in the generated recipe list.

The virtual recipes used in this section are distinct from the logical tools used for virtual extrusion. Both mechanisms use slicer-level tool identifiers, but their interpretations differ (Fig.~\ref{fig:slicer_project_mechanisms}): halftoning projects use virtual recipes to control filament mixtures, while virtual extrusion uses logical tools to trigger process-state changes.


\section{Results}
The results evaluate the compiler as a bridge between heterogeneous design intent and slicer-native fabrication workflows. We first show that scalar attributes can control slicer settings through generated settings meshes, allowing the downstream slicer to plan different toolpaths across an object without manual region assignment. We then evaluate virtual extrusion as a mechanism for process-state control, using calibrated foaming-filament models to translate density and Shore-hardness fields into temperature and flow-rate assignments. Next, we combine slicer-setting and process-state controls in a single compiled project to test whether the same partitioning framework can coordinate multiple fabrication effects. We then evaluate color and material halftoning compilation by converting material-fraction and sRGB fields into ColorMix and FullSpectrum-compatible projects and measuring how the downstream slicing workflows reproduce color gradients. Finally, we quantify the reduction in user interactions obtained by emitting configured \texttt{.3mf} projects directly.

\subsection{Slicer-Setting Control with Generated Settings Meshes}
\label{subsec:results_slicer_setting_control}

First, we evaluate the translation to settings meshes: whether the compiler can translate scalar heterogeneous attributes into slicer-planning controls rather than encoded geometry. In these examples, the source design contains an infill density or fuzzy-skin attribute field. The compiler partitions the field, extracts labeled sub-meshes, and writes a single PrusaSlicer-compatible \texttt{.3mf} project in which each region carries the corresponding slicer setting. We slice each project in PrusaSlicer~2.9.6 and print the examples on a Prusa Core One+ using PLA, a \(0.4~\mathrm{mm}\) nozzle, and \(0.2~\mathrm{mm}\) layers. For the infill examples, we disable top and bottom skins so that the printed artifacts expose the internal infill structure.

\subsubsection{Infill Density}
\label{subsubsec:results_infill_density}

\begin{figure}
    \centering
    \includegraphics[width=1.0\linewidth]{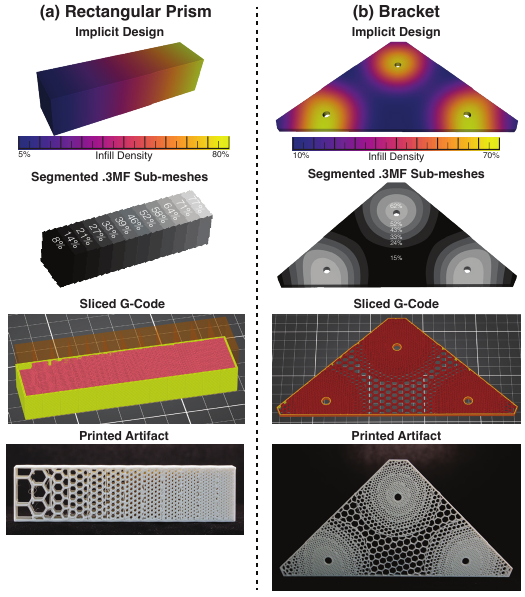}
    \caption{Generated settings meshes for spatially varying infill density. (a) A rectangular prism defines a linear infill-density field from \(5\%\) to \(80\%\), which the compiler partitions into 12 sub-meshes and exports as a configured \texttt{.3mf} project. (b) A bracket defines high infill near three mounting holes and lower infill elsewhere, blended over the object and partitioned into six regions. In both examples, PrusaSlicer plans the local infill from the generated region settings.}
    \label{fig:results_infill_density}
\end{figure}

Figure~\ref{fig:results_infill_density} shows two compiled infill-density fields. The rectangular prism defines a linear infill density gradient from \(5\%\) to \(80\%\) along the \(x\)-axis. We set the partition count to \(N=12\), and the compiler discretizes the scalar field into 12 intervals and outputs 12 aligned sub-meshes. PrusaSlicer then plans each region with the assigned infill density. The printed artifact exposes the expected monotonic change in cell density along the designed axis, showing that the slicer interprets the generated region settings as ordinary local planning controls.

\begin{listing}[t]
\caption{Source definition for the linear infill-density bar in Fig.~\ref{fig:results_infill_density}a.}
\label{lst:results_infill_bar}
\begin{minted}{python}
bar_size_x = 100.0
bar_size_y = 25.0
bar_size_z = 25.0

infill_expr = (
    f"5 + 75 * ((x + {bar_size_x / 2}) / {bar_size_x})"
)

bar = pv.RectPrism(
    pv.Vec3(0, 0, 0),
    pv.Vec3(bar_size_x, bar_size_y, bar_size_z),
)
bar.set_attribute(
    pv.DefaultAttributes.INFILL_DENSITY,
    pv.FloatAttribute(infill_expr),
)
\end{minted}
\end{listing}

The bracket example tests the same compilation mechanism on an application-motivated design. The source field assigns high infill density near the three mounting holes and low infill density elsewhere, then blends the transition over a \(13~\mathrm{mm}\) radius to produce a smooth field over the bracket. The implicit design specifies an infill-density field ranging from \(10\%\) to \(70\%\). We set the partition count to \(N=6\), and the compiler discretizes this range into six infill regions. In the sliced preview and printed artifact, the densest infill appears around the mounting holes and decreases away from them. This result demonstrates a practical use of settings meshes: the designer specifies where increased local stiffness is desired, while the downstream slicer still supplies its native infill pattern, toolpath ordering, and printer-specific planning.

\subsubsection{Fuzzy Skin}
\label{subsubsec:results_fuzzy_skin}

\begin{figure}
    \centering
    \includegraphics[width=0.95\columnwidth]{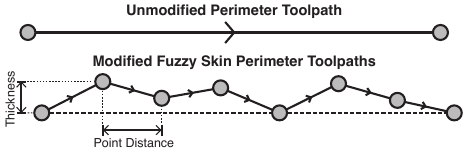}
    \caption{Fuzzy-skin parameters exposed by the downstream slicer. Thickness controls the maximum offset from the nominal perimeter, while point distance controls the spacing between perturbation points along the perimeter.}
    \label{fig:fuzzy_skin_diagram}
\end{figure}

\begin{figure}
    \centering
    \includegraphics[width=0.95\linewidth]{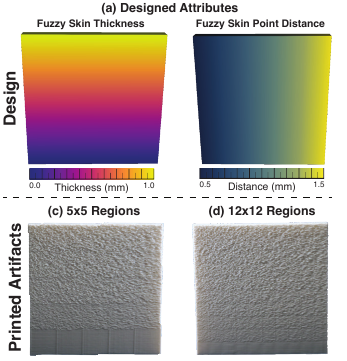}
    \caption{Generated settings meshes for fuzzy-skin parameters. The source design varies fuzzy-skin thickness from \(0.0~\mathrm{mm}\) to \(1.0~\mathrm{mm}\) and fuzzy-skin point distance from \(0.5~\mathrm{mm}\) to \(1.5~\mathrm{mm}\). The printed swatches show that the compiler can assign these slicer settings independently across generated regions.}
    \label{fig:results_fuzzy_skin_prism}
\end{figure}

Figure~\ref{fig:results_fuzzy_skin_prism} visualizes our evaluation of settings meshes for two fuzzy-skin parameters. As shown in Figure~\ref{fig:fuzzy_skin_diagram}, PrusaSlicer exposes two settings to control fuzzy skin: point thickness and point distance. Fuzzy-skin thickness sets the maximum outward perturbation from the nominal perimeter path, while fuzzy-skin point distance sets the spacing between successive perturbation points. Larger thickness increases the amplitude of the surface texture, and larger point distance produces a coarser texture.

The source design assigns fuzzy-skin thickness as a linear field from \(0.0~\mathrm{mm}\) to \(1.0~\mathrm{mm}\) along the vertical axis and fuzzy-skin point distance as a separate linear field from \(0.5~\mathrm{mm}\) to \(1.5~\mathrm{mm}\) along the horizontal axis. The compiler partitions these fields into regular setting regions and writes the corresponding per-region fuzzy-skin overrides. The printed swatches show the expected independent effects: increasing thickness increases roughness amplitude, while increasing point distance produces a coarser and less dense texture. These examples show that settings meshes can control multiple slicer-exposed parameters as spatial gradients without redefining the surface texture as geometry.

\begin{figure}
    \centering
    \includegraphics[width=1.0\linewidth]{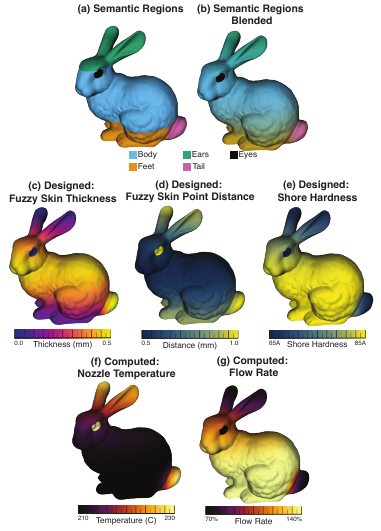}
    \caption{Volumetric semantic attributes for the fuzzy-skin bunny. (a) Semantic regions assign settings to anatomical regions of the bunny. (b) A design-stage blending operation, performed before compilation and slicing, converts the piecewise semantic assignments into smooth volumetric fields. (c,d) The resulting fuzzy-skin thickness and point-distance fields drive settings-mesh compilation. Panels (e--g) define the Shore-hardness, temperature, and flow-rate fields evaluated later for virtual extrusion.}
    \label{fig:results_bunny_attributes}
\end{figure}

We next apply fuzzy-skin settings to a more complex implicit object. Figure~\ref{fig:results_bunny_attributes}a defines semantic regions for the Stanford bunny, including the body, feet, ears, tail, and eyes. These regions are volumetric attributes, not surface-only paint strokes. We highlight this distinction because we use the same implicit volumetric design for this example of surface settings and an example in Sec.~\ref{subsec:results_virtual_extrusion} that modulates shore hardness over the volume. In this subsection, we use only the fuzzy-skin thickness and point-distance fields shown in Fig.~\ref{fig:results_bunny_attributes}c--d. The ears, eyes, and feet receive no fuzzy-skin thickness, while the body and tail receive larger thickness values. The body uses a smaller point distance to create a dense fuzzy texture, and the tail uses a larger point distance to produce a coarser, bushier texture. Table~\ref{tab:bunny_attributes} lists the region values used for both the fuzzy-skin settings and the Shore-hardness field evaluated later.

\begin{table}[t]
\centering
\caption{Semantic attribute values assigned to the bunny. This subsection evaluates the fuzzy-skin settings; Sec.~\ref{subsec:results_virtual_extrusion} evaluates the Shore-hardness field using the same semantic design.}
\label{tab:bunny_attributes}
\small
\setlength{\tabcolsep}{4pt}
\begin{tabular}{@{}lccc@{}}
\toprule
\textbf{Region} &
\textbf{Thickness} &
\textbf{Point dist.} &
\textbf{Shore A} \\
&
\textbf{(mm)} &
\textbf{(mm)} &
\\
\midrule
Feet & 0.0 & 1.0 & 85 \\
Ears & 0.0 & 1.0 & 65 \\
Tail & 0.5 & 1.0 & 65 \\
Eyes & 0.0 & 1.0 & 65 \\
Body & 0.5 & 0.5 & 85 \\
\bottomrule
\end{tabular}
\end{table}

Blending converts the piecewise semantic regions into smooth spatial setting fields, as shown in Fig.~\ref{fig:results_bunny_attributes}b--d. We set the partition count to \(N=12\) for both fuzzy-skin thickness and point distance. The compiler constructs the resulting \(12 \times 12\) joint partition, producing 144 setting combinations before removing empty regions.

\begin{figure*}
    \centering
    \includegraphics[width=0.95\linewidth]{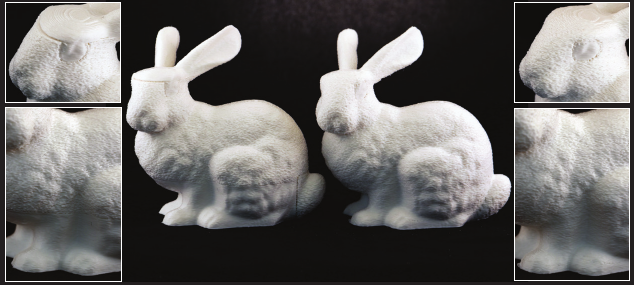}
    \caption{Printed fuzzy-skin bunnies compiled from semantic setting fields. The unblended bunny has sharp texture transitions at semantic-region boundaries, while the blended bunny uses a joint \(12 \times 12\) settings partition to produce smoother texture variation.}
    \label{fig:results_bunny_prints}
\end{figure*}

Figure~\ref{fig:results_bunny_prints} compares printed bunnies compiled with and without blended setting fields. The unblended print has abrupt texture transitions where regions change from fuzzy to smooth. The blended print removes these visible boundary artifacts and produces a gradient change in texture across the body, legs, and tail. This result shows that the compiler can preserve region structure when sharp transitions are desired, or apply blended volumetric attributes when smoother setting variation is required.

These examples establish settings meshes as a direct compilation target for slicer-planning parameters. The compiler does not replace the slicer's infill generator or fuzzy-skin toolpath logic. Instead, it converts heterogeneous attributes into the local settings that the slicer already understands. The next section evaluates the complementary case, where a spatial field controls machine state through virtual extrusion rather than slicer planning.

\subsection{Process-State Control with Virtual Extrusion}
\label{subsec:results_virtual_extrusion}

Next, we evaluate virtual extrusion: whether the compiler can control machine state through slicer-native tool assignments. In this workflow, the compiler partitions process-state fields into discrete regions, maps each region to a logical tool, and relies on the slicer's toolchange events to output the corresponding G-code commands. We demonstrate this mechanism with temperature responsive foaming filaments, where nozzle temperature controls expansion, density, and Shore hardness, and flow-rate compensation preserves geometry. All examples in this section are sliced with PrusaSlicer~2.9.6 and printed by a Prusa MK4S.

\subsubsection{Foaming-Filament Calibration}
\label{subsubsec:results_foaming_calibration}

\begin{figure*}
    \centering
    \includegraphics[width=1.0\linewidth]{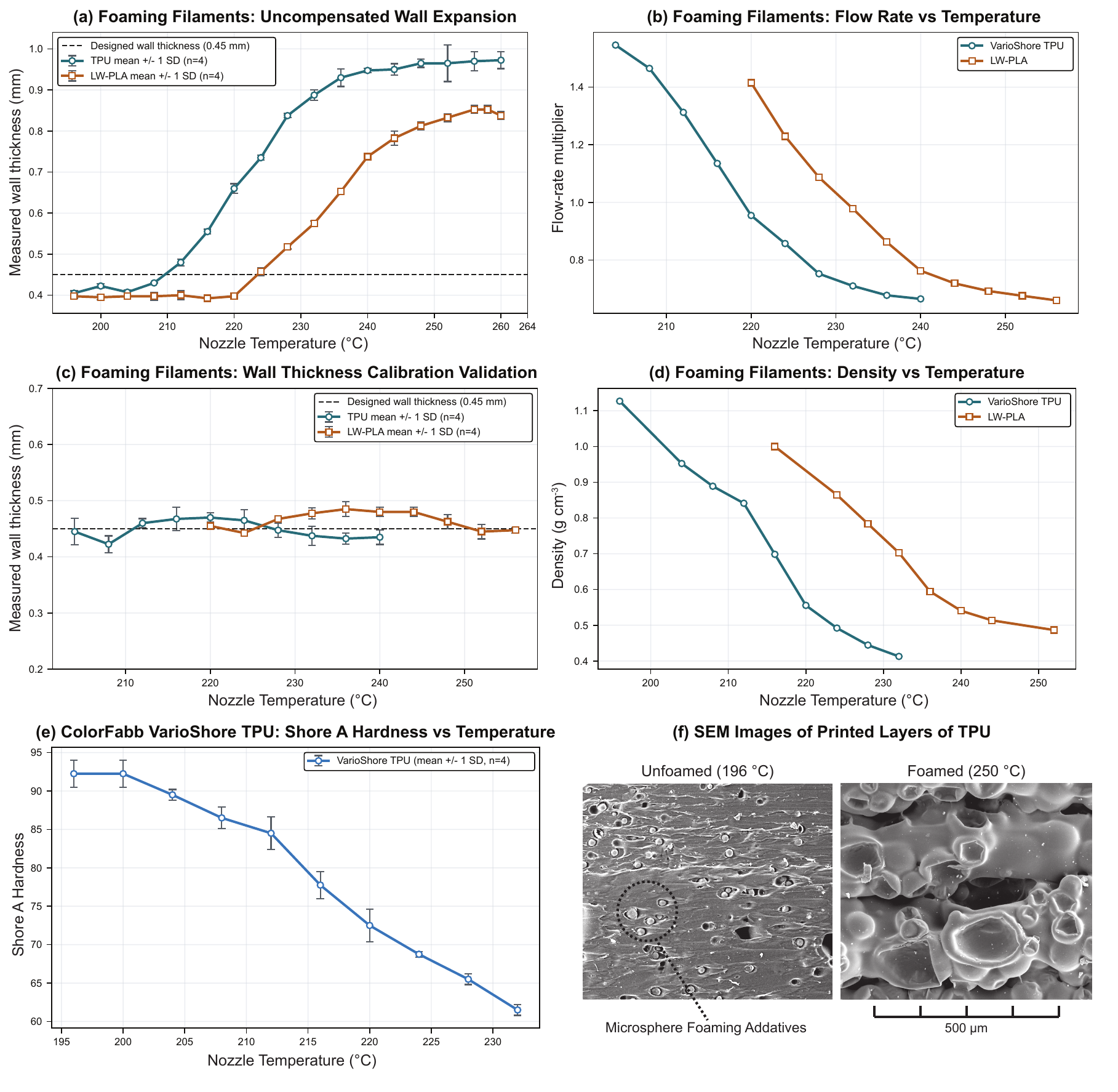}
    \caption{Calibration data for temperature responsive foaming filaments. (a) Uncompensated wall expansion for VarioShore TPU and LW-PLA. (b) Flow-rate multiplier computed from measured wall expansion. (c) Compensated wall-thickness validation, showing measured wall thickness near the \(0.45~\mathrm{mm}\) target across temperature. (d) Density as a function of temperature for compensated solid samples. (e) Shore A hardness as a function of temperature for VarioShore TPU. (f) SEM images of unfoamed and foamed TPU, showing the transition from intact microsphere additives to gas-filled pores.}
    \label{fig:results_foaming_calibration}
\end{figure*}

We first characterize the process models needed to use foaming filaments as controllable attribute-translation targets. We evaluate ColorFabb VarioShore TPU and ColorFabb LW-PLA by printing single-wall calibration samples over a range of nozzle temperatures. Each sample uses a designed wall thickness of \(0.45~\mathrm{mm}\), corresponding to one wall with a \(0.4~\mathrm{mm}\) nozzle, and repeats that wall for \(20~\mathrm{mm}\) in height. For each temperature, we print four samples, measure each wall with a micrometer, and report the mean wall thickness with one standard deviation.


Figure~\ref{fig:results_foaming_calibration}a shows uncompensated wall expansion as a function of nozzle temperature. Both materials expand as temperature increases. VarioShore TPU begins expanding near \(208^\circ\mathrm{C}\) and reaches approximately \(0.97~\mathrm{mm}\) wall thickness at high temperature, corresponding to about \(216\%\) of the designed wall thickness. LW-PLA begins expanding at a higher temperature, near \(222^\circ\mathrm{C}\), and reaches approximately \(0.85~\mathrm{mm}\), or about \(188\%\) of the designed wall thickness. At low temperatures, both materials print slightly below the designed wall thickness, so the compensation model must correct both under-extrusion and over-expansion.

We compute the flow-rate multiplier required at each temperature from the measured wall expansion:
\begin{equation}
    \lambda(T) = \alpha \frac{W_D}{W(T)},
    \label{eq:results_flow_rate_compensation}
\end{equation}
where \(\lambda(T)\) is the flow-rate multiplier at temperature \(T\), \(\alpha=1.0\) is the base flow-rate multiplier, \(W_D=0.45~\mathrm{mm}\) is the designed wall thickness, and \(W(T)\) is the average measured wall thickness at temperature \(T\). Figure~\ref{fig:results_foaming_calibration}b plots the resulting compensation model for both materials. The compiler interpolates the measured calibration data to evaluate \(\lambda(T)\) and converts the resulting multiplier to the percentage form required by the emitted G-code command.

Figure~\ref{fig:results_foaming_calibration}c validates the compensation model. After applying the temperature-dependent flow-rate multiplier, both TPU and LW-PLA remain close to the \(0.45~\mathrm{mm}\) target despite large changes in foaming expansion. This result establishes the process-state pair used by virtual extrusion: temperature controls the material response, while flow-rate compensation preserves bead geometry.

We then use the compensated process to characterize density. We print \(30~\mathrm{mm} \times 30~\mathrm{mm} \times 7~\mathrm{mm}\) solid blocks at \(100\%\) infill with normal top and bottom skins, apply the calibrated flow-rate compensation, measure the mass and external dimensions, and compute density from mass divided by measured external volume. Figure~\ref{fig:results_foaming_calibration}d shows that density decreases monotonically with temperature for both materials. This relationship defines an inverse translation model from target density to nozzle temperature, followed by the temperature-to-flow-rate model in Eq.~\ref{eq:results_flow_rate_compensation}. This calibration defines an inverse translation model: target density maps to nozzle temperature, and temperature then maps to flow-rate compensation.

VarioShore TPU also changes Shore A hardness with temperature. We characterize this relationship by printing \(30~\mathrm{mm} \times 30~\mathrm{mm} \times 7~\mathrm{mm}\) samples in compliance with ASTM D2240 and measuring Shore A hardness with a durometer. Figure~\ref{fig:results_foaming_calibration}e shows that hardness decreases as temperature increases, consistent with the increased foaming and reduced density observed in Fig.~\ref{fig:results_foaming_calibration}d. This calibration defines another inverse translation model: target Shore hardness maps to nozzle temperature, and temperature then maps to flow-rate compensation.

Figure~\ref{fig:results_foaming_calibration}f provides a physical interpretation of these trends. At \(196^\circ\mathrm{C}\), the TPU remains largely unfoamed and the microsphere additives remain visible. At \(250^\circ\mathrm{C}\), the microspheres expand into gas-filled pores within the printed material. This change in pore structure explains why increasing temperature lowers both density and Shore A hardness, and why temperature can serve as the process-state variable for inverse design.

\subsubsection{Density Inverse Design}
\label{subsubsec:results_density_inverse_design}

\begin{figure}
    \centering
    \includegraphics[width=1.0\linewidth]{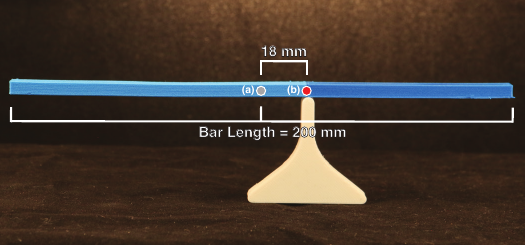}
    \caption{Density inverse design with foaming TPU. The bar is \(200~\mathrm{mm}\) long and is designed to place its center of mass (b) \(18~\mathrm{mm}\) to the right of the geometric center (a). The printed artifact balances at the designed offset, showing that the compiled temperature and flow-rate fields realize the intended density distribution.}
    \label{fig:results_balance_beam}
\end{figure}

Figure~\ref{fig:results_balance_beam} demonstrates density-based inverse design with foaming TPU. The source design is a \(200~\mathrm{mm}\)-long bar whose density field is chosen to place the center of mass \(18~\mathrm{mm}\) to the right of the geometric center. The design uses two calibrated density states from Fig.~\ref{fig:results_foaming_calibration}d: a low-density foamed state of \(0.413~\mathrm{g\,cm^{-3}}\) and a high-density unfoamed state of \(1.123~\mathrm{g\,cm^{-3}}\). The solver selects the density switch point required to achieve the requested offset.

The printed bar balances at the designed off-center location. We determine the realized center of mass by finding the point along the printed artifact where it remains balanced on a support and measuring that point relative to the geometric center. The observed balance point confirms that the density field compiles into a physically meaningful mass distribution. The user specifies the high-level density objective, and the attribute-translation models derive the temperature and flow-rate fields that virtual extrusion applies during slicing and printing.

\subsubsection{Shore-Hardness Inverse Design}
\label{subsubsec:results_shore_hardness_inverse_design}

\begin{figure}
    \centering
    \includegraphics[width=1.0\linewidth]{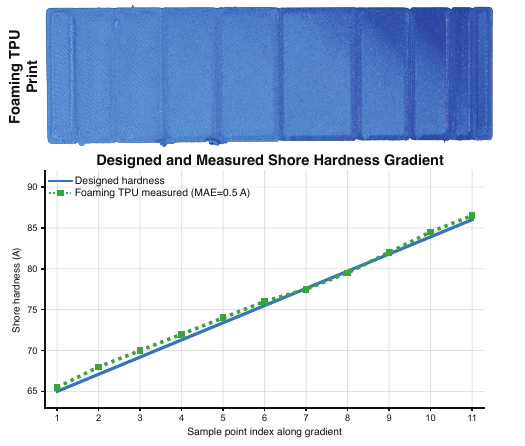}
    \caption{Shore-hardness inverse design with foaming TPU. The source design specifies a linear 65A--85A hardness gradient. The measured values along the printed bar follow the designed gradient with a mean absolute error of \(0.5\) Shore A.}
    \label{fig:results_shore_hardness_validation}
\end{figure}

Figure~\ref{fig:results_shore_hardness_validation} validates the Shore-hardness inverse model from Fig.~\ref{fig:results_foaming_calibration}e on a gradient bar. The source design specifies a linear Shore-hardness field from 65A to 85A. Before compilation, the translation model computes the required temperature field, and the flow-rate compensation model in Eq.~\ref{eq:results_flow_rate_compensation} computes the companion flow-rate field. Virtual extrusion then assigns the resulting process-state tuples to logical tools in a PrusaSlicer project. We measure hardness at 11 evenly spaced points along the printed bar and compare the measurements with the designed values. The measured gradient follows the target field with a mean absolute error of \(0.5\) Shore A, indicating that the compiled process-state assignments preserve the intended mechanical-property gradient over the sampled locations.

\subsubsection{Foaming Fuzzy Bunny}
\label{subsubsec:results_foaming_fuzzy_bunny}

The foaming fuzzy bunny combines process-state control with the settings-mesh control evaluated in Sec.~\ref{subsec:results_slicer_setting_control}. The semantic design is the same one summarized in Table~\ref{tab:bunny_attributes} and visualized in Fig.~\ref{fig:results_bunny_attributes}. The body and feet use the higher hardness value of 85A, while the tail, ears, and eyes use the softer 65A value. The fuzzy-skin thickness and point-distance fields remain active, so the compiler must coordinate two slicer-setting fields with one process-state field.

We set the partition count to \(N=6\) for fuzzy-skin thickness, fuzzy-skin point distance, and nozzle temperature. The compiler then discretizes the three fields into a \(6 \times 6 \times 6\) product space. This example simultaneously exercises multiple compilation targets: a single volumetric design produces a slicer-ready project containing local surface texture settings and local machine-state control. The interaction savings for this joint design are quantified later in Sec.~\ref{subsec:results_interaction_savings}.

\begin{figure}
    \centering
    \includegraphics[width=1.0\linewidth]{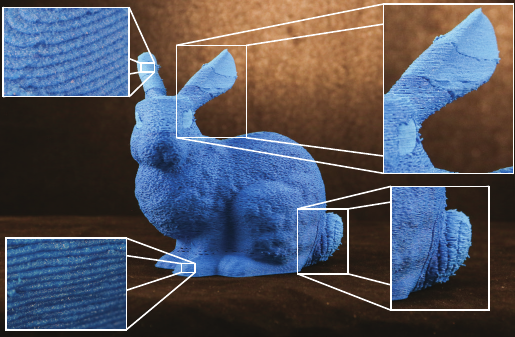}
    \caption{Foaming fuzzy bunny printed from a compiled PrusaSlicer project. The object combines fuzzy-skin settings with virtual-extrusion control of nozzle temperature and flow-rate compensation. Lighter regions correspond to higher-temperature, lower-density, softer foamed TPU, while darker regions correspond to lower-temperature, higher-density, harder TPU.}
    \label{fig:results_foaming_bunny}
\end{figure}

Figure~\ref{fig:results_foaming_bunny} shows the printed artifact. The surface texture follows the fuzzy-skin fields, while the subtle color shift follows the temperature and foaming field. Lighter regions correspond to higher-temperature, lower-density, softer foamed TPU; darker regions correspond to lower-temperature, higher-density, harder TPU. The eye region makes this contrast visible because it preserves a sharp transition. This result shows that virtual extrusion can operate together with settings meshes in a single downstream slicer workflow, allowing high-level volumetric attributes to control both process state and slicer-planned surface texture.

\subsection{Color and Material-Fraction Control with Halftoning Filament Mixing}
\label{subsec:results_color_halftoning}

Finally, we evaluate halftone filament mixing: whether the compiler can translate material-fraction and color attributes into slicer-ready projects for downstream color-mixing workflows. In these examples, the printer is a Prusa XL with five PLA filaments loaded as cyan, magenta, yellow, black, and white (CMYKW). The compiler does not perform the final filament-level halftoning. Instead, it exports ColorMix- and FullSpectrum-compatible \texttt{.3mf} projects containing the virtual recipes and sub-mesh assignments that PrusaSlicer or OrcaSlicer consume during slicing.

\subsubsection{Material-Fraction Gradients}
\label{subsubsec:results_material_fraction_gradients}

We first evaluate direct material-fraction compilation using a \(100~\mathrm{mm} \times 20~\mathrm{mm} \times 20~\mathrm{mm}\) rectangular bar. The source design defines a linear two-material volume-fraction field from \(100\%\) cyan to \(100\%\) magenta along the \(x\)-axis. We set the partition count to \(N=20\), and the compiler discretizes the field into 20 mixture regions before exporting the same abstract material-fraction design to both Prusa ColorMix and Orca FullSpectrum. Both samples are printed on the same Prusa XL with the same CMYKW PLA filaments, a \(0.4~\mathrm{mm}\) nozzle, and \(0.2~\mathrm{mm}\) layers.

\begin{figure}
    \centering
    \includegraphics[width=0.95\linewidth]{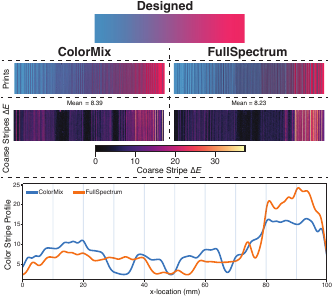}
    \caption{Material-fraction gradient reproduced through downstream halftoning workflows. The source design varies linearly from cyan to magenta and is exported to both Prusa ColorMix and Orca FullSpectrum using the same CMYKW PLA filaments. Coarse-banding analysis shows similar mean banding over the full sample, and both methods produce stronger low-frequency artifacts near the gradient endpoints where one filament has low participation.}
    \label{fig:results_gradient_bar_comparison}
\end{figure}

Figure~\ref{fig:results_gradient_bar_comparison} compares the two printed gradients and analyzes the spatial structure of their halftoning artifacts. This example evaluates how each downstream slicer realizes a prescribed material-fraction field. We therefore focus on the spatial structure of the printed halftone pattern, measuring coarse color banding along the gradient rather than treating the rendered cyan--magenta interpolation as a calibrated color target. Each photograph is converted to CIELAB space, and a local estimate of the intended cyan--magenta gradient is subtracted so that the remaining residual emphasizes halftoning artifacts. We then compute a sliding-window spectral analysis along the \(100~\mathrm{mm}\) sample length. Within each window, the metric integrates residual color variation at long spatial periods orthogonal to the gradient direction. Higher values indicate wider, more visible halftone bands; lower values indicate finer alternation that visually blends more smoothly.

ColorMix and FullSpectrum produce similar mean coarse-banding scores over the full bar, with means of 8.39 and 8.23, respectively. Additionally, the spatial profile shows that both methods produce more visible low-frequency banding near the ends of the gradient, where one material participates at a low fraction. This artifact becomes especially apparent below \(25~\mathrm{mm}\) and above \(75~\mathrm{mm}\), corresponding to mixtures dominated by one filament. The high-magenta end shows stronger visible striping, which is consistent with the observed perceptual dominance of magenta over cyan in these prints. Although the mean scores are similar, FullSpectrum shows a larger coarse-banding peak above \(75~\mathrm{mm}\) than ColorMix. We therefore use this experiment to constrain the full-color palette in the next section: nonzero components in mixed-filament recipes must be at least \(25\%\).

\subsubsection{Full-Color Reproduction}
\label{subsubsec:results_full_color_reproduction}

\begin{figure}
    \centering
    \includegraphics[width=1.0\linewidth]{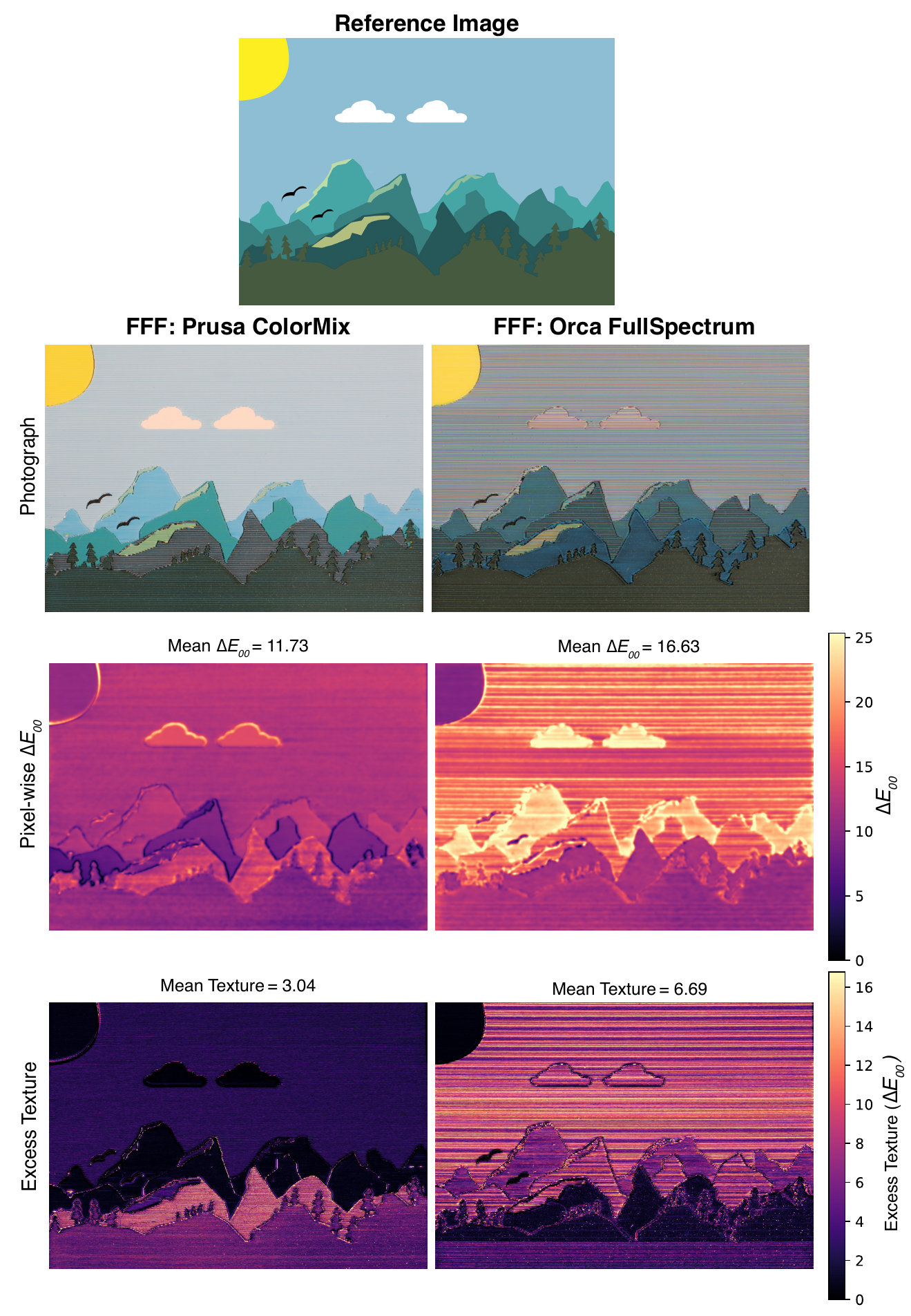}
    \caption{Full-color reproduction from one volumetric RGB color design compiled to two downstream halftoning workflows. Photographs are registered to the reference image and compared using Gaussian-filtered CIEDE2000 color difference, \(\Delta E_{00}\), to measure regional color fidelity. Excess texture isolates high-frequency artifacts introduced by fabrication. For this CMYKW PLA benchmark, ColorMix produces lower mean color error and lower excess texture than FullSpectrum.}
    \label{fig:results_mountain_comparison}
\end{figure}

We next evaluate full-color compilation from an sRGB attribute field. The source design is a mountain-scene image, shown in Figure~\ref{fig:results_mountain_comparison}, imported as an RGB color field and extruded volumetrically into a \(150~\mathrm{mm} \times 110~\mathrm{mm} \times 10~\mathrm{mm}\) object. The same source design is compiled to Prusa ColorMix and Orca FullSpectrum using the same CMYKW PLA base filaments. For both workflows, the palette-selection algorithm receives the same manufacturer-reported color values for the loaded CMYKW base filaments, listed in Appendix~\ref{app:color_palettes}.

The compiler selects a bounded palette from printable recipes containing either a single physical filament or mixtures of two or three base filaments. Mixed recipes must satisfy the \(25\%\) minimum nonzero component fraction motivated by the material-fraction gradient result in Fig.~\ref{fig:results_gradient_bar_comparison}. ColorMix and FullSpectrum use different recipe prediction models, so the optimizer may select different recipes even though both workflows start from the same physical CMYKW filament colors. Appendix~\ref{app:color_palettes} reports the selected palettes for the mountain example.

Figure~\ref{fig:results_mountain_comparison} compares the printed mountain samples. We photograph each sample under fixed camera settings, lighting, and alignment, register the photograph to the digital reference, and compare the images in CIELAB space using CIEDE2000 color difference, \(\Delta E_{00}\). Lower \(\Delta E_{00}\) indicates closer color reproduction. Before computing \(\Delta E_{00}\), we Gaussian filter both the reference and printed sample images in CIELAB space with \(\sigma=12\) pixels. This filtering emphasizes regional color reproduction rather than bead-scale or pixel-scale variation.

We also compute an excess texture metric to isolate fabrication artifacts. First, we remove the smooth color component from both the printed image and the reference image, leaving only fine-scale residual variation. We then compare these residuals and subtract the residual variation already present in the reference. Higher values therefore indicate additional high-frequency texture introduced by the printing process.

ColorMix reproduces the mountain benchmark more accurately than FullSpectrum under these conditions. The ColorMix sample has a mean \(\Delta E_{00}\) of 11.73, compared with 16.63 for FullSpectrum. ColorMix also produces a smoother printed image, with mean excess texture of 3.04 compared with 6.69 for FullSpectrum. The error and texture maps in Fig.~\ref{fig:results_mountain_comparison} show that FullSpectrum introduces stronger horizontal banding across large regions of the image, while ColorMix preserves smoother sky and mountain regions. These results show that, for this CMYKW PLA palette and benchmark image, the compiler can target both slicer workflows from the same volumetric color design, and the downstream workflow choice produces measurable differences in color fidelity and texture quality.

\subsection{Reduction in Manual Slicer Interactions}
\label{subsec:results_interaction_savings}

The preceding examples use the slicer interface as a compilation target rather than as a manual authoring environment. We quantify this difference by counting the number of repetitive slicer interactions that a user would need to recreate the same assignments manually. We define one interaction as a single click or one grouped text-entry action. The counts assume PrusaSlicer~2.9.6 and a favorable manual baseline: the model is already partitioned into the required sub-meshes, all sub-meshes are already aligned, and the user only needs to apply the correct settings or material assignments. The count therefore excludes mesh generation, file export, import, alignment, repair, naming, and slicing. It measures only the repetitive assignment work that the compiler replaces.

Table~\ref{tab:manual_interaction_tasks} lists the primitive interaction counts used in this estimate. Setting infill density requires four interactions per sub-mesh, setting both fuzzy-skin parameters requires nine interactions per sub-mesh, and assigning a material or logical tool requires three interactions per sub-mesh. Our compiler writes these assignments directly into the \texttt{.3mf} project, so the corresponding manual assignment count is zero for the compiled workflow.

\begin{table}[t]
\centering
\caption{Primitive manual interaction counts used to estimate repetitive slicer assignment effort in PrusaSlicer~2.9.6.}
\label{tab:manual_interaction_tasks}
\small
\setlength{\tabcolsep}{5pt}
\begin{tabular}{@{}clc@{}}
\toprule
\textbf{ID} & \textbf{Manual assignment task} & \textbf{Interactions} \\
\midrule
1 & Set both fuzzy-skin parameters & 9 \\
2 & Set infill density & 4 \\
3 & Assign material or logical tool & 3 \\
\bottomrule
\end{tabular}
\end{table}

Table~\ref{tab:manual_interaction_examples} applies these primitive counts to the examples in this paper. For joint partitions, the reported totals assume the worst case in which every cell of the Cartesian-product partition is non-empty and corresponds to a unique settings combination, such that each region produces a separate sub-mesh. The infill-density bar in Fig.~\ref{fig:results_infill_density}a contains 12 sub-meshes, each requiring an infill-density assignment, for a total of 48 manual interactions. The fuzzy-skin bunny uses a \(12 \times 12\) joint partition over two fuzzy-skin parameters, yielding at most 144 sub-meshes and 1,296 manual interactions. The foaming fuzzy bunny combines fuzzy-skin settings with virtual-extrusion tool assignments over a \(6 \times 6 \times 6\) joint partition. Under the same worst-case assumption, recreating this project manually would require both fuzzy-skin assignment and logical-tool assignment for each of 216 sub-meshes, for a total of 2,592 interactions.

\begin{table}[t]
\centering
\caption{Estimated manual slicer interactions required to recreate the assignments used in the paper examples. The compiled workflow embeds these assignments directly in the generated \texttt{.3mf} project.}
\label{tab:manual_interaction_examples}
\small
\setlength{\tabcolsep}{4pt}
\begin{tabular}{@{}lccc@{}}
\toprule
\textbf{Example} & \textbf{Meshes} & \textbf{Task IDs} & \textbf{Interactions} \\
\midrule
Infill bar & 12 & 2 & 48 \\
Infill bracket & 6 & 2 & 24 \\
Fuzzy sheet, \(5 \times 5\) & 25 & 1 & 225 \\
Fuzzy sheet, \(10 \times 10\) & 100 & 1 & 900 \\
Fuzzy bunny & 144 & 1 & 1,296 \\
Foaming fuzzy bunny & 216 & 1, 3 & 2,592 \\
Volume-fraction gradient & 20 & 3 & 60 \\
Mountains & 10 & 3 & 30 \\
\bottomrule
\end{tabular}
\end{table}

These counts show the scaling problem that motivates slicer project compilation. Manual assignment effort grows linearly with the number of generated regions and with the number of independent controls assigned to each region. This estimate further assumes that the user has already completed the non-trivial tasks of partitioning the model and maintaining exact alignment among the resulting regions. A manual workflow may also introduce unnecessary or inconsistent partitions, whereas the compiler constructs the partition directly from the user-specified discretization parameters. By embedding settings, material assignments, and virtual-tool assignments during compilation, the proposed workflow removes this repetitive interface work while preserving the downstream slicer's native planning and preview capabilities.

\subsection{Runtime Benchmarking}
\label{subsec:results_runtime_benchmarking}

We report runtime benchmarking to evaluate whether slicer project compilation remains practical as partition count increases. We use a \(200~\mathrm{mm} \times 50~\mathrm{mm} \times 50~\mathrm{mm}\) rectangular prism and compile three single-gradient benchmarks. The settings-mesh benchmark applies a grid infill-density gradient from \(5\%\) to \(95\%\) along the bar. The virtual-extrusion benchmark applies a temperature gradient from \(200^\circ\mathrm{C}\) to \(250^\circ\mathrm{C}\), with flow-rate compensation computed from Eq.~\ref{eq:results_flow_rate_compensation}. The half-toning benchmark applies a broad sRGB color gradient and exports color-recipe assignments. For each case, we vary the number of actual exported partitions, record compiler wall-clock time, and slice the generated \texttt{.3mf} project in PrusaSlicer~2.9.6 for a Prusa XL.

\begin{figure*}
    \centering
    \includegraphics[width=1.0\linewidth]{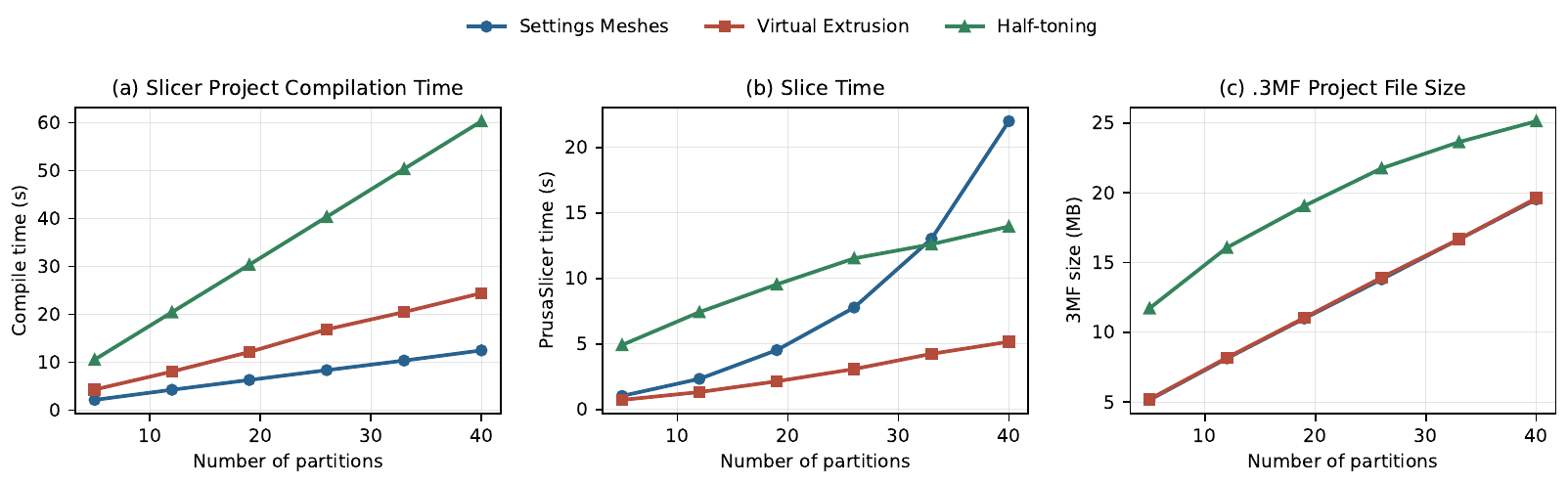}
    \caption{Runtime and storage scaling for slicer project compilation. A \(200~\mathrm{mm} \times 50~\mathrm{mm} \times 50~\mathrm{mm}\) prism is compiled with increasing numbers of exported partitions for grid infill-density settings meshes, temperature-driven virtual extrusion with flow-rate compensation, and color-recipe half-toning. (a) Compiler time grows approximately linearly. (b) PrusaSlicer~2.9.6 slicing time remains below \(21~\mathrm{s}\), although grid infill-density settings meshes grow fastest. (c) Project size increases approximately linearly and remains below \(25~\mathrm{MB}\).}
    \label{fig:benchmarking}
\end{figure*}

Figure~\ref{fig:benchmarking}a shows that compiler time increases approximately linearly for all three targets. Half-toning has the largest constant factor because it serializes color-recipe metadata, while settings meshes and virtual extrusion remain lower-cost. At 40 partitions, the largest compile time is about \(60~\mathrm{s}\), and the other two targets remain near \(20~\mathrm{s}\). Downstream slicing also remains tractable (Fig.~\ref{fig:benchmarking}b). Virtual extrusion and half-toning scale close to linearly, while grid infill-density settings meshes grow faster, likely because many local infill regions increase toolpath-planning complexity in PrusaSlicer. Even so, the longest slicing time is approximately \(21~\mathrm{s}\).

Project size follows the same trend (Fig.~\ref{fig:benchmarking}c). Settings meshes and virtual extrusion have nearly identical linear storage costs, while half-toning produces larger files because each partition carries color-recipe assignments. The largest project remains below \(25~\mathrm{MB}\). These results show that increasing partition count, and therefore gradient fidelity, does not produce prohibitive compile times, slicing times, or project sizes over the tested range.

\section{Conclusion}
We present slicer project compilation as an automated bridge between heterogeneous attribute fields and slicer-native fabrication workflows. Rather than requiring a designer to manually partition a graded object, export aligned meshes, assign region settings, configure material recipes, or post-process G-code, the compiler emits configured \texttt{.3mf} projects that conventional slicers can load, preview, slice, and fabricate. The method reduces continuous attributes to labeled sub-mesh regions and serializes those regions as local slicer settings, process-state assignments, or filament-mixture recipes. Across the examples, the same compilation framework drove slicer-planned infill and fuzzy-skin variation, virtual-extrusion control of nozzle temperature and flow rate, and color or material-fraction assignments for downstream half-toning workflows.

The results also demonstrate that slicer project compilation becomes more useful when paired with calibrated translation models. For temperature responsive foaming filaments, we characterized the relationships among nozzle temperature, flow-rate compensation, density, and Shore A hardness for TPU and PLA workflows. These models allow a designer to specify high-level property fields, such as density or Shore hardness, while the compiler receives the lower-level temperature and flow-rate fields required for fabrication. The density-shifted balance beam, Shore-hardness gradient, and foaming fuzzy bunny show how these translations can be used within a single automated workflow for heterogeneous FFF fabrication. In this workflow, high-level design intent, calibrated material-process behavior, and slicer-ready project generation remain separate steps, but they operate together without manual slicer assignment.

Future work could focus on addressing the following limitations. Continuous fields are approximated by a finite number of project regions, so higher spatial fidelity increases project size and downstream slicer workload. The calibrated process models depend on the material, printer, and print profile, and therefore must be remeasured when the fabrication setup changes. The generated projects also depend on slicer-specific \texttt{.3mf} dialects and on the process-control mechanisms exposed by each slicer. These constraints motivate extensions to additional slicer backends and, more broadly, slicers that can natively accept and plan over true heterogeneous gradients, including continuous toolpath-aware transitions.

By compiling heterogeneous attributes into slicer-ready projects, the system makes field-based design usable within existing FFF workflows while preserving the path planning, preview, support generation, and printer-profile infrastructure of mature slicers. The open-source release provides a reusable implementation compatible with the OpenVCAD design representation, including project-generation backends and calibrated foaming-filament translation models. We intend this release to support further work by the community on functionally graded design, material-process characterization, and automated fabrication workflows in which high-level spatial intent can be carried through to printable artifacts without manual slicer reconstruction.

\begin{acks}
This material is based upon work supported by the Charles Stark Draper Laboratory, Inc. under Contract No. N00030-24-C-6001. Any opinions, findings and conclusions or recommendations expressed in this material are those of the author(s) and do not necessarily reflect the views of Strategic Systems Programs.
\end{acks}


\appendix
\section{Appendix}
\subsection{Code Availability}
The implementation, example projects, calibration data, and scripts used to
produce the figures are available at: 
\url{https://github.com/MacCurdyLab/OpenVCAD-Public}

\subsection{Color Palette Recipes}
\label{app:color_palettes}

This appendix reports the filament colors and selected virtual recipes used for the full-color mountain benchmark in Sec.~\ref{subsubsec:results_full_color_reproduction}. Table~\ref{tab:mountains-base-filaments} lists the physical CMYKW PLA filaments loaded in the printer. Table~\ref{tab:mountains-selected-palettes} lists the bounded palettes selected for the ColorMix and FullSpectrum exports. Both workflows use the same physical filament set, but their recipe prediction models differ, so the selected virtual mixtures are not identical.

\definecolor{mountainbase1}{HTML}{0082AD}
\definecolor{mountainbase2}{HTML}{D10F4F}
\definecolor{mountainbase3}{HTML}{EFD00B}
\definecolor{mountainbase4}{HTML}{3D3E3D}
\definecolor{mountainbase5}{HTML}{E6EAEF}

\begin{table}[h!]
\centering
\caption{Base CMYKW PLA filaments loaded for both ColorMix and FullSpectrum export. The colors supplied to the palette-selection algorithm are the manufacturer-reported hex values for the physical filaments.}
\label{tab:mountains-base-filaments}
\begin{tabular}{@{}r c c l@{}}
\toprule
Slot & Color & Hex & Description \\
\midrule
E1 & \cellcolor{mountainbase1}\phantom{0000} & \texttt{\#0082AD} & cyan \\
E2 & \cellcolor{mountainbase2}\phantom{0000} & \texttt{\#D10F4F} & magenta \\
E3 & \cellcolor{mountainbase3}\phantom{0000} & \texttt{\#EFD00B} & yellow \\
E4 & \cellcolor{mountainbase4}\phantom{0000} & \texttt{\#3D3E3D} & near black \\
E5 & \cellcolor{mountainbase5}\phantom{0000} & \texttt{\#E6EAEF} & white \\
\bottomrule
\end{tabular}
\end{table}

\definecolor{cm0}{HTML}{8DBED3}
\definecolor{cm1}{HTML}{45593F}
\definecolor{cm2}{HTML}{E2E64E}
\definecolor{cm3}{HTML}{397F7F}
\definecolor{cm4}{HTML}{FDFEFE}
\definecolor{cm5}{HTML}{255757}
\definecolor{cm6}{HTML}{47A6A6}
\definecolor{cm7}{HTML}{A1C89D}
\definecolor{cm8}{HTML}{FFF200}
\definecolor{cm9}{HTML}{010101}

\definecolor{fs0}{HTML}{8DBED3}
\definecolor{fs1}{HTML}{45593F}
\definecolor{fs2}{HTML}{397F7F}
\definecolor{fs3}{HTML}{FFF200}
\definecolor{fs4}{HTML}{255757}
\definecolor{fs5}{HTML}{B1BF7F}
\definecolor{fs6}{HTML}{FFFFFF}
\definecolor{fs7}{HTML}{BBDBA5}
\definecolor{fs8}{HTML}{47A6A6}
\definecolor{fs9}{HTML}{91BE9B}

\begin{table*}[h!]
\centering
\caption{Selected palette entries for the mountains example. Swatches show the selected target color; recipes list the physical-filament fractions used to reproduce that color. ColorMix and FullSpectrum use the same physical CMYKW PLA base filaments but different recipe prediction models, so the selected palettes differ.}
\label{tab:mountains-selected-palettes}
\begin{tabular}{@{}c c l@{\qquad}c c l@{}}
\toprule
\multicolumn{3}{c}{\textbf{ColorMix}} &
\multicolumn{3}{c}{\textbf{FullSpectrum}} \\
\cmidrule(lr){1-3}\cmidrule(l){4-6}
ID & Target & Recipe & ID & Target & Recipe \\
\midrule
6  & \cellcolor{cm0}\texttt{\#8DBED3} & E5 75\%, E1 25\% &
6  & \cellcolor{fs0}\texttt{\#8DBED3} & E5 75\%, E1 25\% \\
7  & \cellcolor{cm1}\texttt{\#45593F} & E4 50\%, E1 25\%, E3 25\% &
7  & \cellcolor{fs1}\texttt{\#45593F} & E1 50\%, E2 25\%, E3 25\% \\
8  & \cellcolor{cm2}\texttt{\#E2E64E} & E3 50\%, E5 50\% &
8  & \cellcolor{fs2}\texttt{\#397F7F} & E4 50\%, E1 25\%, E5 25\% \\
9  & \cellcolor{cm3}\texttt{\#397F7F} & E1 75\%, E3 25\% &
9  & \cellcolor{fs3}\texttt{\#FFF200} & E3 75\%, E5 25\% \\
5  & \cellcolor{cm4}\texttt{\#FDFEFE} & E5 100\% &
10 & \cellcolor{fs4}\texttt{\#255757} & E4 75\%, E1 25\% \\
10 & \cellcolor{cm5}\texttt{\#255757} & E4 50\%, E1 25\%, E5 25\% &
11 & \cellcolor{fs5}\texttt{\#B1BF7F} & E5 50\%, E3 25\%, E4 25\% \\
11 & \cellcolor{cm6}\texttt{\#47A6A6} & E1 50\%, E5 50\% &
5  & \cellcolor{fs6}\texttt{\#FFFFFF} & E5 100\% \\
12 & \cellcolor{cm7}\texttt{\#A1C89D} & E5 50\%, E1 25\%, E3 25\% &
12 & \cellcolor{fs7}\texttt{\#BBDBA5} & E5 50\%, E1 25\%, E3 25\% \\
13 & \cellcolor{cm8}\texttt{\#FFF200} & E3 75\%, E5 25\% &
13 & \cellcolor{fs8}\texttt{\#47A6A6} & E1 50\%, E5 50\% \\
4  & \cellcolor{cm9}\textcolor{white}{\texttt{\#010101}} & E4 100\% &
14 & \cellcolor{fs9}\texttt{\#91BE9B} & E1 50\%, E3 25\%, E5 25\% \\
\bottomrule
\end{tabular}
\end{table*}

\bibliographystyle{ACM-Reference-Format}
\bibliography{references, other_sources}

\end{document}